\documentclass[acmsmall]{acmart}
\acmJournal{TODAES}

\AtBeginDocument{%
  }

\usepackage{siunitx}
\usepackage{subcaption}
\usepackage{makecell}

\begin{document}

\title{RFAmpDesigner: A Two-Level LLM-guided Optimization Framework for Automated Differential CMOS RF Amplifier Sizing}

\title[RFAmpDesigner: An Optimization Framework for Automated Differential CMOS RF Amplifier Sizing]{RFAmpDesigner: A Two-Level LLM-guided Optimization Framework for Automated Differential CMOS RF Amplifier Sizing}

\author{Hang Lu}
\orcid{0009-0006-0298-0665}
\affiliation{%
  \institution{Zhejiang University}
  \city{Hangzhou}
  \state{Zhejiang}
  \country{China}}
\email{luhang@zju.edu.cn}

\author{Guochang Li}
\affiliation{%
  \institution{Zhejiang University}
  \city{Hangzhou}
  \state{Zhejiang}
  \country{China}}
\email{gcli@zju.edu.cn}

\author{Qianyu~Chen}
\affiliation{%
  \institution{Zhejiang University}
  \city{Hangzhou}
  \state{Zhejiang}
  \country{China}}
\email{qianychen@zju.edu.cn}

\author{Huiyan~Gao}
\affiliation{%
  \institution{Zhejiang University}
  \city{Hangzhou}
  \state{Zhejiang}
  \country{China}}
\email{11834025@zju.edu.cn}

\author{Shaogang~Wang}
\affiliation{%
  \institution{Zhejiang University}
  \city{Hangzhou}
  \state{Zhejiang}
  \country{China}}
\email{wangsg@zju.edu.cn}

\author{Xuanyu~He}
\affiliation{%
  \institution{Zhejiang University}
  \city{Hangzhou}
  \state{Zhejiang}
  \country{China}}
\email{22334029@zju.edu.cn}

\author{Yiwei~Liu}
\affiliation{%
  \institution{Zhejiang University}
  \city{Hangzhou}
  \state{Zhejiang}
  \country{China}}
\email{3190101962@zju.edu.cn}

\author{Zesen~Chen}
\affiliation{%
  \institution{Zhejiang University}
  \city{Hangzhou}
  \state{Zhejiang}
  \country{China}}
\email{12134010@zju.edu.cn}

\author{Gaopeng~Chen}
\affiliation{%
  \institution{Zhejiang University}
  \city{Hangzhou}
  \state{Zhejiang}
  \country{China}}
\email{gaopeng.chen@raytrontek.com}

\author{Nayu~Li}
\affiliation{%
  \institution{Zhejiang University}
  \city{Hangzhou}
  \state{Zhejiang}
  \country{China}}
\email{linayu@zju.edu.cn}

\author{Chunyi~Song}
\affiliation{%
  \institution{Zhejiang University}
  \city{Hangzhou}
  \state{Zhejiang}
  \country{China}}
\email{cysong@zju.edu.cn}

\author{Xiaokang~Qi}
\correspondingauthor
\authornote{
Xiaokang Qi is the corresponding author.
\\
This work was supported by the National Natural Science Foundation of China, Key Project No.~62434008, by the National Key Research and Development Program of China under Grant No.~2023YFB4403304, supported by the Fundamental Research Funds for the Central Universities 226-2026-00029.
}
\affiliation{%
  \institution{Zhejiang University}
  \city{Hangzhou}
  \state{Zhejiang}
  \country{China}}
\email{qixk@zju.edu.cn}

\author{Zhiwei~Xu}
\affiliation{%
  \institution{Zhejiang University}
  \city{Hangzhou}
  \state{Zhejiang}
  \country{China}}
\email{xuzw@zju.edu.cn}


\renewcommand{\shortauthors}{H. Lu et al.}

\begin{abstract}
  Automated RF amplifier sizing is challenging because the design process is inherently hierarchical, the parameter space is high-dimensional and strongly coupled, and high-fidelity circuit simulation is expensive. Conventional optimization and reinforcement learning methods typically require many iterations and make limited use of RF domain knowledge, making them costly under tight simulation budgets. Large language models (LLMs) provide broad prior knowledge and reasoning capabilities, but are unreliable for direct numerical optimization. This paper presents RFAmpDesigner, a two-level optimization framework for automated schematic-level sizing of differential CMOS RF amplifiers. At the outer level, the circuit-parameter space is abstracted into a resource-allocation space, where LLMs make high-level allocation decisions and adjust them according to feedback from the inner level. At the inner level, RF design knowledge is used to decompose the flat sizing problem into a sequence of low-dimensional subproblems. Conventional optimizers then solve these subproblems to find circuit implementations consistent with the assigned resources. This structure allows numerical optimizers to focus on low-dimensional searches in the circuit-parameter space, while the LLM reasons and acts over a physically meaningful design space using simulation feedback. The main evaluation covers low-noise amplifier (LNA), driver amplifier (DA), and power amplifier (PA) designs within the target differential CMOS RF amplifier family. Experimental results show that it achieves a higher success rate under a limited simulation budget in both easy and hard regimes.
\end{abstract}




\maketitle

\section{Introduction}
EDA tools have continuously driven the semiconductor industry toward higher performance and greater design efficiency. Analog circuit automation, however, remains far less mature than its digital counterpart. Analog designs are governed by continuous parameter spaces and lack the discrete abstractions that enable digital synthesis. RF design, as a branch of analog, introduces further difficulty. The design task is inherently hierarchical, requiring the circuit to first operate in the correct frequency band before meeting in-band performance targets. Moreover, the design parameters are strongly coupled. These characteristics make automated RF sizing a particularly demanding open problem.

\begin{figure}[!t]
\centering
\includegraphics[width=0.6\columnwidth]{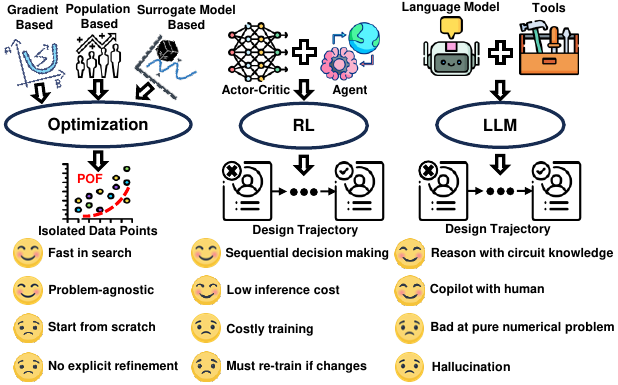}
\caption{Conceptual overview of widely adopted methodologies for circuits design automation.}
\label{fig1}
\end{figure}

Existing approaches to analog circuit parameter sizing fall into two broad categories: optimization-based methods\cite{DE, BO, NSGABU1, NSGABU2, bottomup_19, topdown} and AI-based methods. The latter can be further divided into reinforcement learning\cite{RL,GCNRL,RLAnalog1} and large language model approaches\cite{LADAC, Analog_BO_LLM, LEDRO, AmpAgent_fo, AnaFlow, AnalogCoder, LLMUSO, AutoSizer, AmpAgent,ASTRA,Artisan,Atelier, AnalogCoder-Pro}. Fig.~\ref{fig1} summarizes the strengths and limitations of all three. Classical optimization methods are fast and largely problem-agnostic. When applied in a flat search\cite{DE, BO}, they require a substantial simulation budget to converge. Hierarchical variants\cite{NSGABU1, NSGABU2, bottomup_19, topdown} have been used mainly at the system level and depend on precomputed macro-models such as Pareto-optimal fronts. In either case, they do not explicitly interpret RF-specific intermediate failure modes to revise their exploration and exploitation strategies. RL methods are well suited to sequential decision-making and can achieve a low inference cost after training. However, training requires a large number of samples and the learned policy is sensitive to changes in parameter space. In RF front-end design, where the number of cascade stages often varies across tasks, this can lead to a substantial training simulation budget. 

Existing LLM-based methods use circuit knowledge and simulation feedback to propose sizing candidates\cite{AnaFlow, LADAC}, sometimes in conjunction with Bayesian optimization\cite{LLMUSO, Analog_BO_LLM}. Other methods guide numerical search by narrowing parameter ranges\cite{LEDRO}, extracting sizing relationships\cite{AmpAgent_fo}, or adapting search spaces and optimization strategies\cite{AutoSizer}. Physics-guided methods use intermediate quantities, such as stage transconductance, compensation capacitance, and $g_m/I_D$, to guide transistor sizing\cite{AmpAgent,ASTRA,Artisan,Atelier}, with some also supporting topology selection or modification. Related work also explores circuit generation\cite{AnalogCoder} and joint topology and parameter optimization\cite{AnalogCoder-Pro,ChaeSurvey}. Although hierarchical decomposition and parameter-subset selection can reduce search complexity, methods that retain device-level decision interfaces still require the LLM to reason about parameter values and search ranges. In RF amplifiers, the coupled effects of device and matching-network parameters on system performance are difficult to infer directly. This can limit LLM-guided search when the decision representation leaves cross-stage resource trade-offs implicit.

The above limitations point to a mismatch between the design representation and the capabilities of existing solvers. Numerical optimizers are effective at low-dimensional parameter search, but become inefficient when system-level resource trade-offs must be discovered in a strongly coupled joint space. LLMs can reason about such trade-offs from feedback, but are unreliable at direct numerical sizing. To address this gap, this paper presents RFAmpDesigner, a two-level optimization framework for CMOS RF amplifier sizing, to separate these two roles. At the outer level, the LLM coordinates stage-wise resource decisions, including current, gain, noise, and linearity budgets. RF-aware tool middleware translates each decision into tractable circuit-level sizing problems at the inner level. The resulting feedback indicates whether an allocation is feasible and how it should be revised. In this way, the optimization loop follows the hierarchy of RF amplifier design instead of treating all circuit parameters as a single flat vector. Our contributions can be summarized as follows:

\begin{enumerate}
\item RFAmpDesigner reformulates specification-driven, schematic-level RF amplifier sizing as a hierarchical nested optimization problem. The outer level searches over a compact resource vector with explicit physical meaning. The inner level decomposes the resource-to-parameter mapping into sequential low-dimensional subproblems using domain knowledge.

\item At the outer level, this work introduces an LLM-based search agent that reasons, decides, and revises within the resource coordinate system. A retrieval-augmented memory of previously explored design points reduces redundant exploration.

\item At the inner level, this work introduces a resource-to-parameter compiler that bridges resource decisions and circuit-level implementation. RF cascade constraints, a pre-characterized device database, and local optimization are combined to solve each subproblem in a principled order.

\item Experiments on LNA, DA, and PA benchmarks demonstrate that the proposed framework achieves higher schematic-level success rates under a limited simulation budget in both easy and hard regimes, with more consistent convergence in the former and better sampling efficiency in the latter. Post-EM validation of selected designs further assesses the extent to which schematic-level performance is retained after component-level EM tuning.
\end{enumerate}

The rest of this paper is organized as follows. Section~\ref{sec2} presents the formulation and design principles. Implementation details of RFAmpDesigner are introduced in Section~\ref{sec3}. Section~\ref{sec4} presents the experimental results. Section~\ref{sec5} concludes the paper.

\section{Preliminaries}
\label{sec2}
\subsection{Problem Statement}
Let \(\mathbf{x}\in\mathcal{X}\subset\mathbb{R}^{n}\) denote the full set of circuit design variables for a multi-stage RF amplifier chain. Given a simulator-based evaluation function \(\mathcal{E}(\cdot)\), the sizing task is

\begin{equation}
\begin{aligned}
\mathbf{x}^{\star}
&=
\arg\max_{\mathbf{x}\in\mathcal{X}}
F\big(\mathcal{E}(\mathbf{x})\big) \\
\text{s.t.}\quad
& g_i\big(\mathcal{E}(\mathbf{x})\big)\leq 0
\quad i=1,\ldots,M \\
\end{aligned}
\label{eq:flat_optimization}
\end{equation}
where \(F(\cdot)\) is a scalarized figure-of-merit or reward function, and \(g_i(\cdot)\) represents performance constraints such as gain, bandwidth, noise figure, input/output matching, linearity, stability, and power consumption. 
The formulation above describes the general sizing problem. This work addresses sizing problem for a prescribed topology within the target circuit family introduced in Section 2.2.

Flat optimization over the joint space faces three additional difficulties in the RF setting:
\begin{enumerate}
\item Full-chain simulation of an RF front-end is expensive per evaluation, so the optimizer has fewer iterations within a practical compute budget and less room to converge.
\item Before gain, noise, and linearity can be meaningfully optimized, the passive network must first place the response in the target band. This prerequisite makes feasible solutions sparse in the joint space.
\item Frequency response is tightly coupled to low-level component values such as inductances and capacitances, so the distance from a random initial point to the feasible region varies widely across seeds, leading to high variance in convergence cost.
\end{enumerate}

\subsection{Target Circuit Family and Structural Properties}
Hierarchical decomposition is the standard remedy: surrogate models of individual blocks reduce the effective search dimension at the system level, after which each block is sized independently~\cite{Rutenbar2007Hierarchical}. The approach is general but requires offline model construction, and composed accuracy degrades when inter-block coupling invalidates the assumed boundary conditions.

This work targets differential neutralized common-source, transformer-coupled multi-stage RF amplifier chains, a widely used architecture in recent mm-wave RF front-ends. Among ISSCC LNA and PA designs published between 2022 and 2026 (30 surveyed), 83\% employ neutralized differential common-source stages cascaded through magnetically coupled resonator (MCR) interstage networks. Recent automated sizing efforts target the same structure~\cite{NSGABU1, competitor_B}.

Three structural properties of this family make direct hierarchical decomposition feasible without constructing a complete block-level behavioral surrogate for every stage:
 
\begin{enumerate}
\item \textbf{Reverse isolation.}
The neutralization capacitor suppresses the Miller feedback and improves reverse isolation. As a result, the port impedances of each stage can be approximated by extracted parallel RC networks, which allows adjacent resonators to be designed nearly independently.
 
\item \textbf{Analytical solvability of the matching networks.}
Input, interstage, and output matching networks take classical L, T, or $\pi$ topologies. Once the source and load impedances are specified, these networks admit analytical or semi-analytical initialization, followed by low-dimensional local tuning to account for finite Q, parasitics, and bandwidth constraints.
 
\item \textbf{Cascade-law governed performance.}
Treating each gate-to-gate interval as one stage, the chain's overall noise figure and $\mathrm{IIP}_3$ follow the Friis equation and the cascade formula, respectively, so system-level specifications decompose directly into per-stage budgets.
\end{enumerate}

\subsection{Hierarchical Reformulation}
RFAmpDesigner reformulates RF amplifier sizing as a hierarchical nested optimization. Directly searching the netlist-level parameter space is both high-dimensional and semantically opaque to an LLM. The outer level therefore lifts the decision space to a compact resource vector over which an LLM can reason with physical intuition. Mapping a resource allocation back to device parameters for a multi-stage chain remains nontrivial. The inner level exploits cascade structure to decompose this mapping into sequential subproblems, each solvable under local constraints.

Instead of searching the physical parameter vector $\mathbf{x}$ directly, the outer optimizer operates on a compact resource vector $\mathbf{r}$:
\begin{equation}
\mathbf{r}
=
[
\boldsymbol{\rho}_{P},
\boldsymbol{\rho}_{G},
h_{\mathrm{NF}},
h_{\mathrm{lin}}
]
\label{eq:resource_vector}
\end{equation}
where $\boldsymbol{\rho}_{P}$ is the stage-wise current allocation, $\boldsymbol{\rho}_{G}$ is the stage-wise gain allocation, $h_{\mathrm{NF}}$ and $h_{\mathrm{lin}}$ are the noise and linearity headrooms. Each coordinate of $\mathbf{r}$ corresponds to a design decision that an RF engineer makes when distributing limited resources across cascaded stages.

To make the resource abstraction executable, this work introduces a resource-to-parameter compiler. For a given resource decision $\mathbf{r}$, target specification $\mathbf{s}$, and prepared design knowledge $\mathcal{D}$, the compiler maps the resource-level decision to a concrete circuit parameter vector:
\begin{equation}
(\mathbf{x},\boldsymbol{\delta})
=
T_{\mathcal{T}}(\mathbf{r};\mathcal{D},\mathbf{s})
\label{eq:compiler_mapping}
\end{equation}
where $\mathcal{T}$ denotes the selected topology template.
The compiler is required to check two groups of conditions:
\begin{equation}
\begin{aligned}
\mathbf{v}_{\mathrm{spec}}(\mathbf{x};\mathbf{s}) &\leq \mathbf{0},\\
\mathbf{v}_{\mathrm{res}}(\mathbf{x};\mathbf{r}) &\leq \boldsymbol{\epsilon}_{r}
\end{aligned}
\label{eq:compiler_conditions}
\end{equation}
The first condition enforces the specification-derived requirements, while the second condition enforces consistency with the resource decision. If all checks in (\ref{eq:compiler_conditions}) are satisfied, $\boldsymbol{\delta}=\mathrm{success}$. Otherwise, $\boldsymbol{\delta}$ reports the failure reason and the violated metric to the outer search loop.
In practice, $T_{\mathcal{T}}(\cdot)$ is not evaluated by solving one monolithic optimization problem. It is implemented as a Gauss-Seidel-style ordered block synthesis process. Let $\mathbf{z}_k$ denote the local variables of the $k$-th synthesis block. Given the resource decision $\mathbf{r}$ and target specification $\mathbf{s}$, the blocks are solved sequentially as
\begin{equation}
\mathbf{z}_k^{\star}
=
\arg\min_{\mathbf{z}_k\in\mathcal{Z}_k(\mathbf{z}_{<k}^{\star})}
C_k
\left(
\mathbf{z}_k;
\mathbf{z}_{<k}^{\star},
\mathbf{r},
\mathbf{s},
\mathcal{D}
\right),
\quad k=1,\ldots,K
\label{eq:ordered_block_synthesis}
\end{equation}
where $\mathbf{z}_{<k}^{\star}$ denotes the solutions of the preceding blocks.
The complete circuit parameter vector is then assembled by
\begin{equation}
\mathbf{x}
=
\mathcal{A}_{\mathcal{T}}
\left(
\mathbf{z}_1^{\star},\ldots,\mathbf{z}_K^{\star}
\right)
\label{eq:block_assembly}
\end{equation}

In RFAmpDesigner, the ordered subproblems correspond to active sizing and biasing, critical-stage impedance matching, frequency bandwidth planning and full-chain evaluation. Implementation details are introduced in Section~\ref{sec3}.

With this reformulation, the aforementioned three difficulties are relaxed. This structure reduces wasted full-chain simulations because arbitrary low-level points are no longer evaluated directly. If any block fails to meet its local feasibility condition, the compiler terminates early and returns the corresponding diagnostic flag without invoking full-chain simulation. The passive network is also synthesized inside the inner loop to place the response near the target band, so gain, noise, and linearity are optimized on meaningful candidates rather than on random off-band responses. Moreover, the seed-dependent search over low-level L/C values is replaced by small-scale sub-optimizations, which reduces the variance in convergence cost.

\subsection{Agentic Search over Resource Space}
After the resource-to-parameter compiler is introduced, the remaining search problem is no longer a direct netlist-level sizing task. The outer optimizer chooses resource variables and interprets diagnostic feedback returned by the compiler (insufficient noise headroom, excessive linearity burden, infeasible passive bandwidth). This interface is suitable for LLM-based agents because both the decisions and the failure signals are expressed in RF design concepts. The inner optimization is still assigned to conventional algorithms such as particle swarm optimization~(PSO).

The agent maintains a compact search state
\begin{equation}
\mathbf{z}_t =
\left\{
\mathbf{s},
\mathcal{T},
\mathcal{Q}_t,
\mathbf{r}^{\mathrm{best}}_t,
\mathbf{m}^{\mathrm{best}}_t,
d_t,
\mathcal{H}_t
\right\}
\label{eq:agent_state}
\end{equation}
where $\mathbf{s}$ is the target specification, $\mathcal{T}$ is the topology template, $\mathcal{Q}_t$ is the candidate queue, $\mathbf{r}^{\mathrm{best}}_t$ and $\mathbf{m}^{\mathrm{best}}_t$ denote the best resource decision and its simulated performance found so far, $d_t$ is the diagnostic flag returned by the compiler, and $\mathcal{H}_t$ stores the previous attempts.
At each iteration, the agent selects one of four resource-level actions:
\begin{equation}
a_t \in
\left\{
\mathsf{Search},
\mathsf{Retrieve},
\mathsf{Prune},
\mathsf{Refine}
\right\}
\label{eq:agent_action}
\end{equation}
The $\mathsf{Search}$ action proposes new resource allocations and invokes the compiler to synthesize circuit candidates. The $\mathsf{Retrieve}$ action reuses previous resource decisions or failure patterns under similar specifications. The $\mathsf{Prune}$ action removes candidates that are unlikely to satisfy the specifications. The $\mathsf{Refine}$ action locally adjusts a promising resource allocation according to the observed specification violation.

The agent is not used as a numerical optimizer, a task for which LLMs are poorly suited. Its role is to reason about design trade-offs, make decisions, and revise its strategy from feedback. Unlike the scalarized-feedback optimization baselines, the LLM controller receives structured RF diagnostic feedback and can associate different failure modes with different resource-level actions. For example, a first-stage noise violation requires a different resource adjustment from an output-linearity violation.

\begin{figure}[!t]
\centering
\includegraphics[width=\columnwidth]{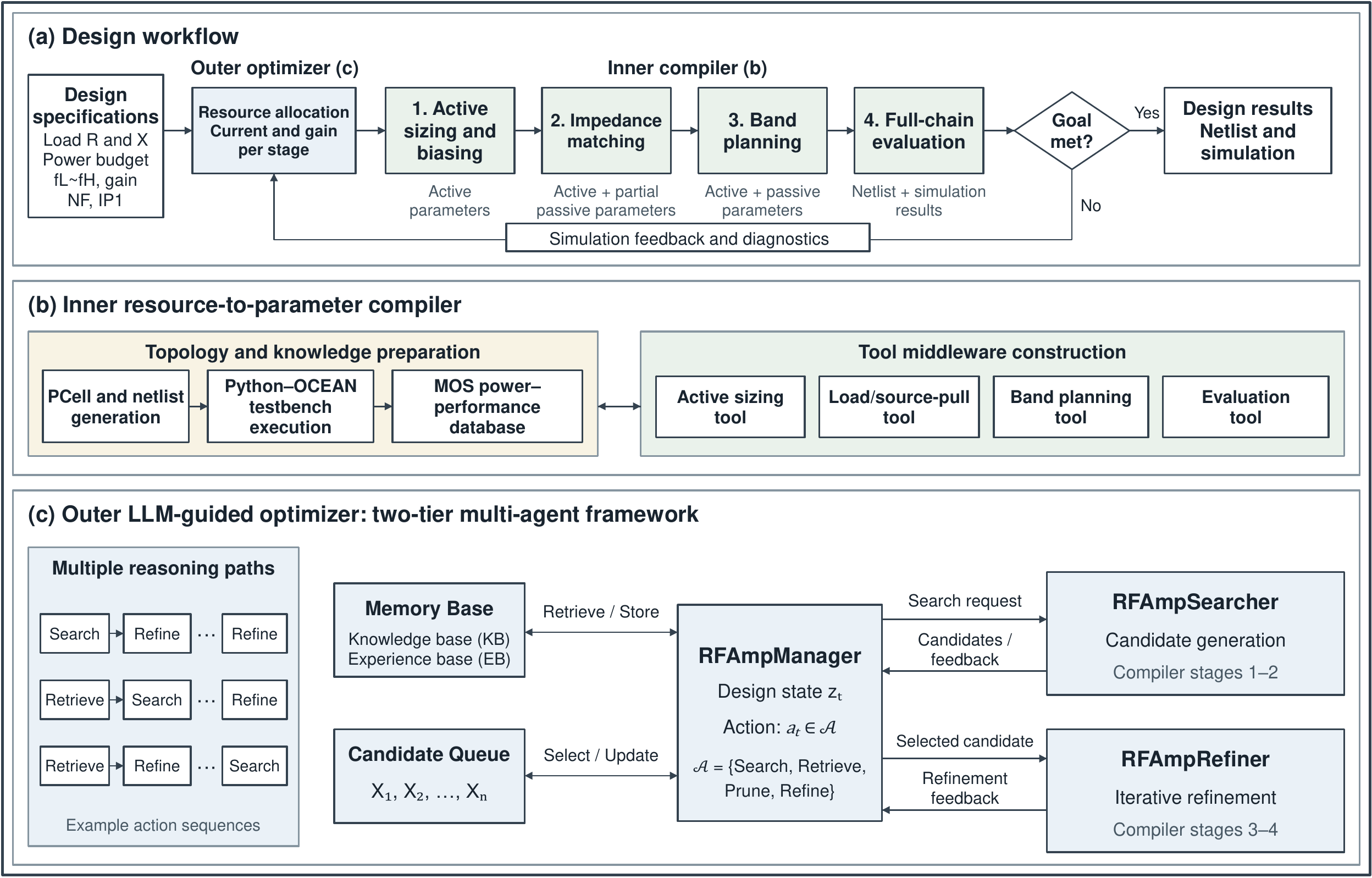}
\caption{(a) The proposed two-level optimization framework. (b) The inner resource-to-parameter compiler of RFAmpDesigner consisting of Topology and knowledge preparation (yellow block) and multi-fidelity tool middleware (green block). (c) The outer LLM-guided optimizer of RFAmpDesigner (blue block): Agents collaborate to decide the resource vector (current and gain distribution of each stage) iteratively.}
\label{fig2}
\vspace{-14pt}
\end{figure}

\section{Proposed Framework}
\label{sec3}
Fig.~\ref{fig2} presents an overview of RFAmpDesigner. Given a set of user specifications, the LLM agent interprets the requirements and makes outer-level decisions, producing a resource allocation over current and gain across stages. The inner-level compiler then takes this resource vector, invokes a sequence of tools to solve the resulting subproblems, and returns a concrete circuit implementation along with simulation results. If the design fails to meet specifications, the LLM examines the compiler's feedback, reasons about the failure, and revises its allocation. This loop repeats until the task is completed or the evaluation budget is exhausted. This section presents the framework following the top-to-bottom and left-to-right flow of Fig.~\ref{fig2}: the prepared knowledge base $\mathcal{D}$, the tool definition for each subproblem within the compiler $T_{\mathcal{T}}$, and the implementation details of the outer-level agent.

\subsection{Topology and Knowledge Preparation}
This subsection comprises two parts. The first is building automation scripts for circuit evaluation, including Cadence parameterized cell generators for active and passive devices, shell scripts for parasitic extraction and EM simulation, testbench netlist and OCEAN templates for full-chain evaluation, and a top-level shell script with configurable parameters. The second is constructing lookup tables that map low-level circuit parameters to performance metrics such as power consumption. Leveraging the reverse isolation property of the target circuit family discussed earlier, automation scripts are used to sweep across device sizes and bias conditions, building the lookup table automatically:
\begin{equation}
\mathcal{D}_{\mathrm{act}}
=
\left\{
(W,V_{\mathrm{GS}},I_{\mathrm{dc}},g_m,
Z_{\mathrm{in}},Z_{\mathrm{out}})
\right\}
\label{eq:active_database}
\end{equation}
The final dataset encompasses cascode devices with widths ranging from \SI{45}{\micro\meter} to \SI{180}{\micro\meter} in \SI{9}{\micro\meter} increments, differential common-source devices from \SI{45}{\micro\meter} to \SI{243}{\micro\meter} in \SI{9}{\micro\meter} steps and bias voltages from \SI{300}{\milli\volt} to \SI{500}{\milli\volt} in \SI{25}{\milli\volt} increments. The above grids comprise 351 device-size and bias combinations. Generating each lookup-table entry requires one DC simulation and three S-parameter simulations, taking approximately 5 s on our workstation. The estimated sequential simulation time for constructing the lookup table is therefore 1,755 s, or approximately 30 min. This estimate covers the device-characterization simulations and excludes script development and the preceding parasitic extraction. Within the same CMOS process, the resulting table can be reused across sizing tasks and circuit topologies that use the characterized device configurations within the covered operating range.

\subsection{Inner Compiler: Multi-fidelity tool middleware}
This subsection describes how the resource vector $\mathbf{r}$ is translated into a concrete circuit implementation $\mathbf{x}$ through four sequential subproblem-solving steps.

\subsubsection{Stage 1}\textbf{Active Device Sizing and Bias Selection}

With the lookup table $\mathcal{D}_{\mathrm{act}}$, finding device sizes and bias conditions for a given current is straightforward. However, for a given current, multiple realizations may exist, and different choices reflect different design priorities. A larger device with a smaller overdrive yields a higher $g_m/I_D$ ratio and therefore higher transconductance, at the expense of larger $C_{gs}$. From a small-signal perspective, this is a power-efficient strategy that extracts more gain from the same current. Higher $g_m$ also improves noise performance, and biasing near the threshold voltage may introduce gain expansion, which benefits linearity. On the other hand, from a large-signal perspective, larger $C_{gs}$ makes it harder for the preceding stage to deliver sufficient drive, and the lower power gain penalizes output power. Yet biasing near threshold can provide gain expansion, which aids large-signal drive. To balance these competing considerations, only the largest and smallest device sizes are retained for each current budget. Gain stages are partitioned into critical and non-critical groups, so the same current allocation gives rise to four size realizations in total, ensuring that the search covers the boundary of the design trade-off space.

\subsubsection{Stage 2}\textbf{Impedance Matching for Critical Performance Stages}

LNA and PA serve different functions and operate in distinct regimes. An LNA works under small-signal conditions, where the first stage dominates noise performance. A PA works under large-signal conditions, where the last stage determines output power and efficiency. For these critical stages, source-pull and load-pull characterization is necessary to approach the optimum device performance. Because the LNA and PA critical stages employ different active device structures in our experiments, their implementation details are discussed separately below.

The LNA employs a single-ended cascode with inductive source degeneration. Because the reverse isolation of this topology is limited, load-pull is required in addition to source-pull. In practice, the source inductor $L_s$ has a finite quality factor, so the degeneration inductor itself introduces a noise resistance into the signal path. The real part of the input impedance contributed by feedback is $g_mL_s/C_{gs}$. The closer this value is to $50~\Omega$, the larger $L_s$ must be, and the more noise it contributes. For this reason, $L_s$ is often not sized to provide the full $50~\Omega$ match on its own. Instead, an additional impedance transformation network is inserted between the input pad and the transistor gate. Although this network introduces its own loss, that loss is not necessarily worse than the noise penalty of a larger $50~\Omega$, creating room for optimization. Here, an L-type input matching network is adopted.
Combining these observations with standard noise theory yields the noise figure expression:
\begin{equation}
\begin{aligned}
F_{\min}&=1+\frac{2}{\sqrt{5}}\frac{\omega C_{gs}}{g_{m}}\sqrt{\gamma\delta\left(1-\left|c\right|^{2}\right)},\\
F_{add}&=(\frac{\gamma}{\alpha g_m R_s})(1-\frac{Q_{opt}}{Q_s})^2+F_{L-MN},\\
F&=F_{min}+F_{add}
\end{aligned}
\label{eq:active_database}
\end{equation}
Inspecting this formula reveals a useful decomposition: $C_{gs}/g_m$ of the device sets the minimum noise figure $F_{min}$, while the additional noise contribution $F_{add}$ is determined once the source impedance $R_s$ is chosen. $R_s$ itself depends on $L_s$ and the load impedance. Once $R_s$ is fixed, the L matching network parameters and the resulting noise $F_{L-MN}$ can be computed analytically from impedance transformation relations.
This analysis motivates our optimization strategy for the LNA critical stage. We first fix the device size and bias, then optimize only the load network and the value of $L_s$. For each candidate, the input matching network corresponding to the $R_s$ is computed analytically rather than treated as an optimization variable. The objective is to minimize $F_{add}$. This approach not only reduces the search dimension by removing the input matching network from the optimization, but also makes the noise figure traceable to its root cause. If $F$ cannot be brought within specification, the conclusion is that the device size and bias must be revised.

The PA employs a differential common-source stage with neutralization capacitors. The source impedance is set to the conjugate of the input impedance, only load-pull is performed. In the PA optimization, the compiler again fixes the device size and bias according to the allocated current, then optimizes only the real and imaginary parts of the load impedance seen from the PA drain to find a load that satisfies the output power and PAE requirements. Once the target load is determined, the output matching network is optimized separately. For each candidate, the matching loss is computed analytically according to the matching network topology. $S_{21}$ is obtained through high-fidelity simulation. The objective is to minimize both. As in the LNA case, matching network parameters are moved into a low-cost optimization loop, preserving speed without sacrificing accuracy.
PSO is adopted with a population size of 10, inertia weight w of 0.8, and acceleration coefficients $c_1$ and $c_2$ of 0.5. The ranges of the optimization variables are listed in the left column of Table~\ref{tab12}, and the loss function of LNA is defined in Equation~(\ref{eq:critical_cost1}):
\begin{align}
\mathrm{NF}_{\mathrm{violation}}
&=
\max\!\bigl(0,\,
\mathrm{NF}_{\mathrm{sim}}
-
(\mathrm{NF}_{\mathrm{user}}-h_{\mathrm{NF}})
\bigr)
\notag\\
\mathrm{S}_{11,\mathrm{violation}}
&=
\max\!\bigl(0,\,
\mathrm{S}_{11,\mathrm{sim}}-\mathrm{S}_{11,\mathrm{user}}
\bigr)
\notag\\
G_{\mathrm{violation}}
&=
\max\!\bigl(0,\,
G_{\mathrm{req}}-G_{\mathrm{sim}}
\bigr)
\notag\\
\mathrm{cost}_{\mathrm{LNA}}
&=
1000\,\mathrm{NF}_{\mathrm{violation}}
+
1000\,\mathrm{S}_{11,\mathrm{violation}}
\notag\\
&\quad+
1000\,G_{\mathrm{violation}}
\label{eq:critical_cost1}
\end{align}

The PA loss function follows the same structure, with the target specifications replaced by $P_{out}$, PAE, and $S_{22}$. 
The objective is a threshold violation cost: strict feasibility is achieved only when all violation terms are zero, i.e., \(\mathrm{cost}=0\). The common coefficient 1000 is therefore used only to amplify residual violations rather than to express different metric preferences. In practice, we use \(\mathrm{cost}<100\) as a relaxed stopping criterion, allowing small residual errors before the subsequent full-chain evaluation.

\begin{table}[!t]
\centering
\caption{Variable ranges for Stage 2 and Stage 3 optimization of LNA}
\label{tab12}
\begin{tabular}{cc|cc} 
\hline
\multicolumn{2}{c|}{\textbf{Stage 2}} & \multicolumn{2}{c}{\textbf{Stage 3}} \\
\textbf{Variables} & \textbf{Ranges} & \textbf{Variables} & \textbf{Ranges} \\
\hline
$k$ & $[0.1,0.7]$ & $k$ & $[0.1,0.8]$ \\
$L_1(\text{pH})$ & $[100, 1000]$ & $\omega_0(\text{GHz})$ & $[\omega_L,\omega_H]$ \\
$L_2(\text{pH})$ & $[100, 1000]$ & $Q_0$ & $[0.1,8]$ \\
$R_1(\Omega)$ & $[100, 1000]$ & $C(fF)$ & $[C_2,200]$ \\
$R_2(\Omega)$ & $[100, 1000]$ &  & \\
$C_1(fF)$ & $[0, 50]$ &  & \\
$C_2(fF)$ & $[0, 50]$ &  & \\
$L_s(pH)$ & $[0, 150]$ &  & \\
\hline
\end{tabular}
\end{table}

\begin{figure}[!t]
\centering

\begin{subfigure}{\columnwidth}
    \centering
    \includegraphics[width=0.5\columnwidth]{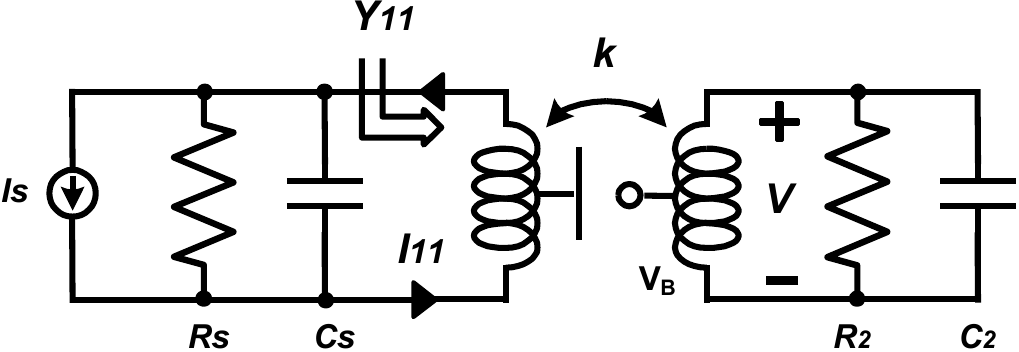}
    \caption{The lumped model of MCR.}
    \label{fig4}
\end{subfigure}

\vspace{2mm}

\begin{subfigure}{\columnwidth}
    \centering
    \includegraphics[width=0.4\columnwidth]{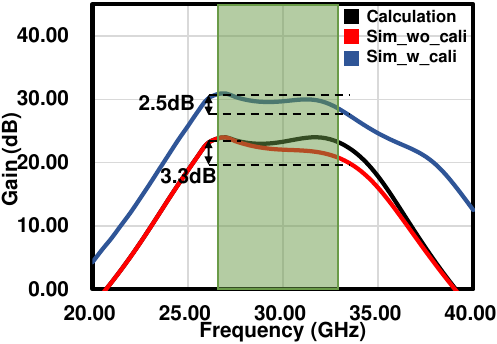}
    \caption{MCR frequency response: calculation vs. simulation
    (with/without calibration).}
    \label{fig5}
\end{subfigure}

\vspace{2mm}

\begin{subfigure}{\columnwidth}
    \centering
    \includegraphics[width=0.4\columnwidth]{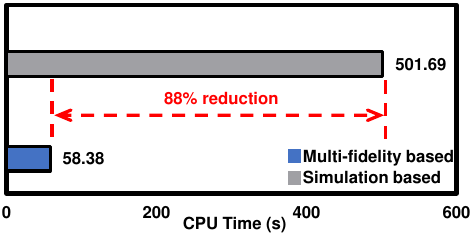}
    \caption{Time consumption: full simulation vs. multi-fidelity.}
    \label{fig6}
\end{subfigure}

\caption{Results of MCR.}
\label{fig:mcr_results}
\end{figure}

\subsubsection{Stage 3}\textbf{Bandwidth Planning}

The next step is establishing frequency selectivity by tuning the resonators to the desired band. This amounts to a large-scale joint optimization over n resonator stages, and direct use of high-fidelity simulation would be prohibitively slow. Using the reverse isolation property introduced earlier, the compiler first calculates the frequency response of each stage shown in Fig.~\ref{fig4} analytically using Equation~(\ref{eq2}), then sums the individual responses to obtain the overall transfer function. 
\begin{align}
Y_{11}(s) &=
\frac{(1-k^2)Q^2 s^4 + 2(1-k^2)Q\omega_0 s^3 + (1-k^2+2Q^2)\omega_0^2 s^2 + 2Q\omega_0^3 s + Q^2 \omega_0^4}
     {(1-k^2)R_{\mathrm{S}} \omega_0 Q s^3 + (1-k^2)R_{\mathrm{S}} \omega_0^2 s^2 + R_{\mathrm{S}} \omega_0^3 Q} \notag \\
R_L &= \frac{1}{\Re\!\bigl(Y_{11}\bigr)}, \qquad
C_L = \frac{\Im\!\bigl(Y_{11}\bigr)}{s} \notag \\
Z_{21}(s) &=
\frac{k Q \omega_0^3 \sqrt{R_{\mathrm{S}} R_{\mathrm{2}}}\, s}
     {(1-k^2)Q^2 s^4 + 2(1-k^2)Q\omega_0 s^3 + (1-k^2+2Q^2)\omega_0^2 s^2 + 2Q\omega_0^3 s + Q^2 \omega_0^4} \notag \\
I_{11}(s) &= g_m \frac{50\,R_{\mathrm{S}} C_{\mathrm{S}} s + R_{\mathrm{S}} + 50}{R_{\mathrm{S}}}\,
            \frac{R_{\mathrm{S}} (R_{\mathrm{L}} C_{\mathrm{L}} s + 1)}{R_{\mathrm{S}} R_{\mathrm{L}} (C_{\mathrm{S}} + C_{\mathrm{L}}) s + R_{\mathrm{S}} + R_{\mathrm{L}}} \notag \\
V(s) &= \frac{I_{11}(s)\, Z_{21}(s)}{2}. \label{eq2}
\end{align}

The analytical model provides low-cost frequency-response evaluations for bandwidth planning. To account for model error and imperfect reverse isolation, the compiler evaluates the candidate through full-chain simulation and uses the residual to correct the analytical response. The optimization and calibration steps may be repeated, with at most five calibration iterations per bandwidth plan. If the constraints remain unmet at this limit, the plan is rejected.
As shown in Fig.~\ref{fig5}, Fig.~\ref{fig5}(b) illustrates this correction process. The black curve is the analytical prediction, and the red curve is the corresponding full-chain simulation before calibration, which exceeds the 3-dB ripple limit. The blue curve is the simulated response after calibration and satisfies that limit. This illustration imposes only the ripple constraint, which allows the calibrated response to have a different absolute gain. Practical optimization also imposes the gain requirement. Fig.~\ref{fig5}(c) reports CPU times of 501.69 s for full-simulation evaluation and 58.38 s for multi-fidelity evaluation. The result reported as 58.38 s required no more than three calibration iterations.
The optimization uses simulated annealing with default parameters and exploits the symmetry of MCR. The variables and their ranges are listed in the right column of Table~\ref{tab12}.

\subsubsection{Stage 4}\textbf{Full-Chain Performance Evaluation}

After the first three stages, every block of the circuit has been sized through its respective sub-optimization. In this final stage, the compiler aggregates all results and runs high-fidelity simulation to produce accurate performance metrics for each concrete circuit implementation $\mathbf{x}$. In addition to the user-specified targets, this stage reports per-stage noise and linearity breakdowns, providing the agent with the diagnostic detail needed to guide its next decision.

\subsection{Multi-agent collaboration}
For most straightforward tasks, a single compiler invocation suffices. Difficult tasks that require tight trade-offs demand more invocations to explore the design space, and a conventional optimizer or RL agent can handle this role. However, under a strict time budget or simulation limit, the optimizer must search in a coarse-to-fine manner, terminate unpromising explorations early, and reuse information from previous trials to guide subsequent ones. This workflow is widely adopted in academic practice~\cite{HuaWangLNASurvey,HuaWangPASurvey}. Inspired by this, RFAmpDesigner realizes this policy with a two-tier multi-agent structure.

\begin{figure}[!t]
\centering
\includegraphics[width=0.95\columnwidth]{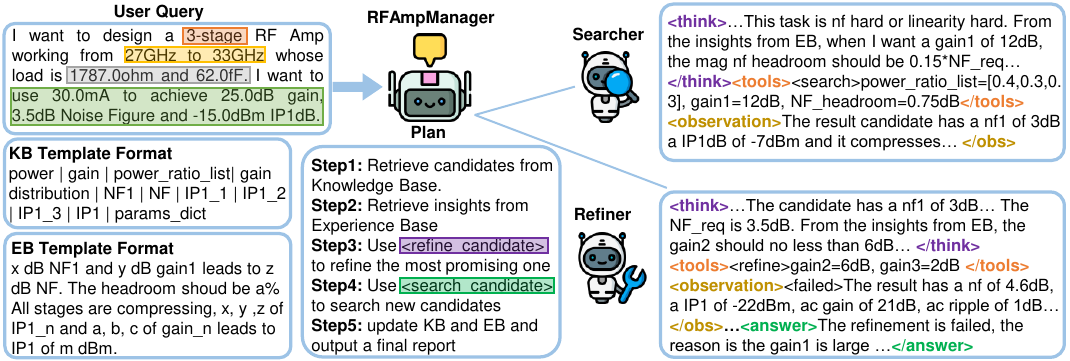}
\caption{Proposed multi-agent framework and template formats of each part.}
\label{fig7}
\end{figure}

RFAmpDesigner consists of three agents that collaborate to carry out coarse-to-fine search. RFAmpManager serves as the main agent, maintaining and updating the design state $\mathbf{z}_t$ and scheduling actions. Two separate-context subagents, RFAmpSearcher and RFAmpRefiner, handle different portions of the compiler pipeline. RFAmpSearcher executes stages 1 and 2 to produce candidate designs. RFAmpRefiner takes a given candidate, executes stages 3 and 4, and iterates on them based on feedback. This division is motivated by cost: both LNA and PA critical stages require high-fidelity evaluation, making stages 1 and 2 computationally expensive. Stages 3 and 4, by contrast, use the mixed-fidelity scheme described earlier and are cheaper. Decoupling the costly evaluation of critical components from the faster refinement stage avoids repeated high-cost evaluations and improves end-to-end efficiency. Every agent executes a ReAct-based~\cite{yao2022react} multi-turn loop that interleaves reasoning with tool calls under a standardized schema. The conversation formats are illustrated in Fig.~\ref{fig7}.

To enable early termination of unpromising searches and reuse of prior exploration, a memory mechanism is introduced in two forms: a Knowledge Base (KB) and an Experience Buffer (EB). Each time the compiler is fully executed, the resulting full-chain parameters and measured performance are stored as structured records in the KB, serving as accumulated design knowledge that guides future resource allocation. Intermediate results, including per-stage noise, gain, and linearity, are stored in the EB, providing partial evidence that allows unpromising candidates to be pruned before a full simulation completes.
Upon receiving a new user query, RFAmpManager can issue a $\mathsf{Retrieve}$ action that queries the KB for the top-k design points most similar to the target specifications. The retrieved points are preloaded into the candidate queue as priors for refinement. The $\mathsf{Prune}$ action retrieves designs from the EB whose critical-stage intermediate results are close to the current candidate, ranks them, and injects them into the prompt. In this way, the approximate Friis and cascade IIP1 formulas are anchored by actual simulated data. The cascade formulas do not define a fixed pruning rule. Instead, they provide evidence for the LLM to decide whether the candidate should be pruned, deprioritized, or refined through current reallocation, critical-stage re-search, or gain redistribution.

\section{Experiments}
\label{sec4}
This section first describes the experimental setup and evaluation methodology. The proposed two-level optimization framework is then evaluated from both inner and outer optimization perspectives. The inner compiler is validated through comparisons with flat optimization baselines, while the contribution of LLM-guided search is investigated under hard-regime tasks, where simulation budget allocation becomes critical. Additional experiments analyze the impacts of different agent components and evaluate different LLM backbones. Then, the exploitation behavior and figure-of-merit trade-offs are studied. Finally, extended RF design examples with post-EM validation are discussed.

\subsection{Experiment Settings}
\subsubsection{Implementation details} 

All agents are implemented using the chat-completion API with temperature of 0, number of samples of 1 and maximum completion tokens of 1024. The tools of inner compiler invoke Cadence OCEAN scripts via Python–shell interfaces for circuit simulations and employ particle swarm optimization and simulated annealing from \texttt{scikit-opt} to solve sub-optimization tasks. All circuit simulations are performed using Cadence Spectre. Experiments run on a Linux workstation with two AMD EPYC~7763 (64 cores each).

\begin{figure}[!t]
\centering
\includegraphics[width=0.6\columnwidth]{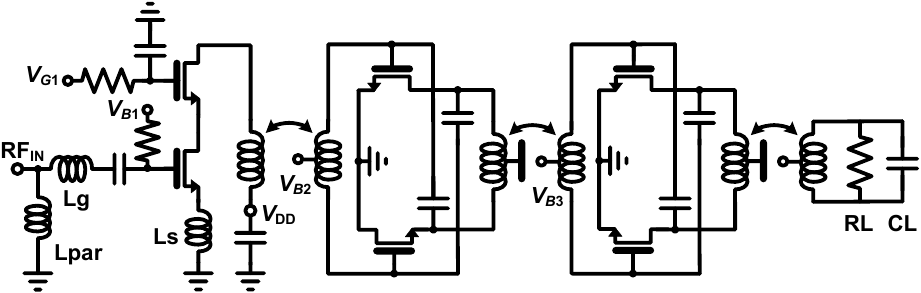}
\caption{Topology used for LNA and DA.}
\label{fig3}
\end{figure}

\begin{table}[!t]
\centering
\caption{The Design Benchmark}
\label{tab3}
\begin{tabular}{|c|c|c|c|c|c|c|c|}
\hline
Specs & Type & $f_{\mathrm{c}}$ (GHz)
& $f_{\mathrm{BW}}$
& Power (mA)
& Gain (dB)
& $\mathrm{NF}$ (dB)
& $\mathrm{IP1dB}$ (dBm) \\
\hline
$S_1$ & LNA & 10 & 20\% & 30 & 25 & $\leq$5 & $\geq$-25 \\
\hline
$S_2$ & LNA & 30 & 20\% & 30 & 25 & $\leq$5 & $\geq$-25 \\
\hline
$S_3$ & LNA & 50 & 20\% & 30 & 20 & $\leq$5 & $\geq$-20 \\
\hline
$S_4$ & LNA & 30 & 10\% & 30 & 25 & $\leq$5 & $\geq$-25 \\
\hline
$S_5$ & LNA & 30 & 60\% & 30 & 25 & $\leq$5 & $\geq$-25 \\
\hline
$S_6$ & LNA & 30 & 80\% & 30 & 20 & $\leq$5 & $\geq$-20 \\
\hline
$S_7$ & LNA & 30 & 20\% & 30 & 25 & $\leq$2.5 & $\geq$-25 \\
\hline
$S_8$ & LNA & 30 & 20\% & 15 & 25 & $\leq$5 & $\geq$-25 \\
\hline
$S_9$ & DA & 30 & 20\% & 30 & 25 & $\leq$5 & $\geq$-15 \\
\hline
$S_{10}$ & DA & 30 & 20\% & 30 & 25 & $\leq$3.5 & $\geq$-15 \\
\hline
\end{tabular}
\end{table}

\subsubsection{Benchmark} 

A custom benchmark of ten tasks is constructed using a single topology (shown in Fig.~\ref{fig3}) that covers both LNA and DA functions, as shown in Table~\ref{tab3}. The target levels were informed by published amplifier designs. Tasks~1–3 fix the fractional bandwidth at 20\% and examine operation from 10 to 50~GHz at 20\% fractional bandwidth. Tasks~4–6 fix the center frequency at 30~GHz and examine fractional bandwidths from 10\% to 80\%. Task~7 tightens the noise figure limit to 2.5 dB. Task~8 imposes a tighter power budget. These tasks probe device and matching-network behavior, bandwidth planning, and cross-stage resource trade-offs within a common topology.
The easy group ($S_1$–$S_8$) uses fixed active stages and a current allocation of [0.4, 0.3, 0.3]. The hard group ($S_9$–$S_{10}$) requires balancing NF and linearity, combining a stricter $\mathrm{IP1dB}$ target of $\geq$ -15 dBm with NF limits of 5 and 3.5 dB, respectively, while allowing joint resource-allocation optimization. The groups use one-hour and five-hour wall-clock limits, respectively, to evaluate compiler decomposition and outer-loop control. These labels describe the experimental setup rather than difficulty for every method.
Prior studies on LNAs with comparable structures ~\cite{benchOP1,benchOP2} report typical OP1dB values around 0 dBm, so tasks 9 and 10 are treated as DA test cases. Since on-chip LNAs and DAs typically drive a transistor gate, the load for all ten tasks is set to gate impedance of a medium-sized active stage ($1787~\Omega \,\parallel\, 62~\mathrm{fF}$).

\begin{table}[t]
  \centering
  \caption{Design variable definitions of different spaces.}
  \label{tab:full_space}
  \setlength{\tabcolsep}{4pt}

  \begin{tabular}{@{}p{0.26\linewidth}p{0.70\linewidth}@{}}
    \hline
    \textbf{Space} & \textbf{Variables} \\
    \hline

    Full Space (30-D) &
    $W_1, W_2, W_3, Len_1, Len_2, Len_3,$
    $V_{GS1}, V_{GS2}, V_{GS3}, W_{CG},$ \newline
    $Cap_2, Cap_3, L_s, L_{\mathrm{par}}, L_g,$
    $k_1, L_1, L_2, R_1, R_2, C_1, C_2,$ \newline
    $k_{\mathrm{2nd}}, \omega_{\mathrm{2nd}},
    Q_{\mathrm{2nd}}, C_{\mathrm{2nd}},$
    $k_{\mathrm{3rd}}, \omega_{\mathrm{3rd}},
    Q_{\mathrm{3rd}}, C_{\mathrm{3rd}}$ \\
    \hline

    Fix Active (18-D) &
    $k_1, L_1, L_2, R_1, R_2, C_1, C_2,$
    $L_s, L_{\mathrm{par}}, L_g,$ \newline
    $k_{\mathrm{2nd}}, \omega_{\mathrm{2nd}},
    Q_{\mathrm{2nd}}, C_{\mathrm{2nd}},$ \newline
    $k_{\mathrm{3rd}}, \omega_{\mathrm{3rd}},
    Q_{\mathrm{3rd}}, C_{\mathrm{3rd}}$ \\
    \hline

    Stage~2 Reduction (16-D) &
    $k_1, L_1, L_2, R_1, R_2, C_1, C_2, L_s,$
    \newline
    $k_{\mathrm{2nd}}, \omega_{\mathrm{2nd}},
    Q_{\mathrm{2nd}}, C_{\mathrm{2nd}},$ \newline
    $k_{\mathrm{3rd}}, \omega_{\mathrm{3rd}},
    Q_{\mathrm{3rd}}, C_{\mathrm{3rd}}$ \\
    \hline

    Resource Space (8-D) &
    $G_1, G_2, G_3, P_1, P_2, P_3,$
    $h_{\mathrm{NF}}, h_{\mathrm{lin}}$ \\
    \hline

  \end{tabular}
\end{table}

\subsubsection{Baselines and metrics} 

On easy tasks(Tasks~1–8), a single compiler invocation is typically sufficient. To demonstrate the advantage of sequential subproblem solving over flat search, the following methods are compared: flat-search vanilla Bayesian Optimization (BO)\cite{BOTool} and Genetic Algorithms (GA)\cite{sko}; BO and GA augmented with the stage~2 tool for input matching network synthesis; and the full compiler. Two LLM-based baselines are also included: AnalogCoder\cite{AnalogCoder}, which uses an LLM to solve the sizing problem directly, and ADO-LLM\cite{Analog_BO_LLM}, which uses an LLM to assist BO. The complete variable dimensions for each method are listed in Table~\ref{tab:full_space}.
On hard tasks (Tasks~9--10), flat-search baselines did not produce feasible solutions within the budget. Therefore, BO, GA, and ADO-LLM are equipped with the compiler and operate in the resource space. To isolate the benefit of adaptive budget allocation from resource-space reduction, BO use fixed-$K$ schedules, where each Search call is followed by $K\in{1,3,5,10}$ Refine calls. BO-Adaptive provides a non-LLM adaptive comparison using the same diagnostic interface. It starts with three Refine calls, extends to five if the best cost decreases by at least 1\% between calls one and three, and extends to ten if the same condition holds between calls three and five. Otherwise, it returns to Search. Its scheduling decisions use this scalar cost-improvement rule, whereas RFAmpDesigner uses the structured diagnostics to revise resource allocations and select subsequent actions.
The hyperparameters of GA and BO are determined by grid search over five random trials on representative tasks 1, 4, 7, and 8, selected based on success rate and the number of evaluations needed to converge. GA uses a population size of 28, 9 generations, and a mutation probability of 0.01. BO uses the UCB acquisition function with $\beta= 2.576$, initialized with 20 samples.
RL methods are not included in the baselines. An internal experiment was conducted to assess their practicality (details in \ref{app:rl}): a PPO agent implemented in Ray was trained on three discrete passive MCR stages, where the overall frequency response is the sum of the individual MCR responses, mimicking the band planning task. The results indicate that roughly 5 million simulation calls are needed for convergence. If the same step count were executed using the 20s full-chain Spectre environment, the training time would exceed 1000 days, which would exceed our available computational budget. So the comparison focuses on optimization-based and LLM-assisted methods that can be evaluated under the same wall-clock constraints.

Three primary metrics are reported: (i) average wall-clock time to a feasible solution, (ii) pass@1 success rate over independent runs (distinct random seeds), and (iii) token consumption for all LLM-based methods to evaluate computational overhead. 
A run is considered successful if the final design achieves cost < 100, allowing a total residual specification violation below 0.1 given the penalty coefficient of 1000. Within each task group, all methods use the same wall-clock budget, feasibility criteria, and full-chain stopping condition.

\subsubsection{Agent backbones}
To evaluate the robustness of RFAmpDesigner across different LLM architectures, we employ a diverse set of backbones ranging from frontier closed-source models to specialized open-source variants. First, four state-of-the-art frontier models are evaluated, including GPT-5.1, Claude-Sonnet-4.5, Gemini-2.5-Flash, and DeepSeek-V3.2 to benchmark the performance and cost-effectiveness. Second, to investigate the impact of model scale and specialized fine-tuning, two models from the Qwen3 family~\cite{qwen3} are adopted: Qwen3-Coder-480B-A35B-Instruct (mid-scale of activated 35B; code-oriented and fine-tuned for tool use) and Qwen3-30B-A3B-Instruct (small-scale of activated 3B; no dedicated tool-use fine-tuning). The latter setting probes the small model’s ability to understand and operationalize complex RF-circuit concepts.

\begin{table}[t]
  \centering
  \caption{Optimized Circuit Parameters for Spec.~5}
  \label{tab:params}

  \renewcommand{\arraystretch}{1.12}
  \setlength{\tabcolsep}{4.5pt}

  \begin{tabular}{@{}c@{\hspace{18pt}}c@{}}

    \begin{tabular}[t]{@{}ll@{}}
      \toprule
      \textbf{Parameter} & \textbf{Value} \\
      \midrule


      $W_1$                 & 72\,$\mu$m \\
      $W_2$                 & 54\,$\mu$m \\
      $W_3$                 & 54\,$\mu$m \\
      $V_{\mathrm{GS},1}$   & 500\,mV \\
      $V_{\mathrm{GS},2}$   & 375\,mV \\
      $V_{\mathrm{GS},3}$   & 375\,mV \\

      \addlinespace[2pt]
      \midrule
      \addlinespace[1pt]


      $L_s$                 & 0\,pH   \\
      $L_g$                 & 93\,pH  \\
      $L_{\mathrm{par}}$    & 155\,pH \\

      \bottomrule
    \end{tabular}

    &

    \begin{tabular}[t]{@{}lccc@{}}
      \toprule
      \textbf{Parameter} & \textbf{1st MCR}
                   & \textbf{2nd MCR}
                   & \textbf{3rd MCR} \\
      \midrule

    $k$                & 0.69 & 0.52 & 0.53 \\
    $L_1$\,(pH)        & 700  & 464 & 100  \\
    $L_2$\,(pH)        & 600  & 464 & 100  \\
    $R_1$\,($\Omega$)  & 1000  & 374  & 76  \\
    $R_2$\,($\Omega$)  & 600  & 483  & 79  \\
    $C_1$\,(fF)        & 10   & 54   & 165   \\
    $C_2$\,(fF)        & 0   & 1    & 103    \\

      \bottomrule
    \end{tabular}

  \end{tabular}
\end{table}

\begin{figure}[t]
  \centering

  \begin{subfigure}[t]{0.3\columnwidth}
    \centering
    \includegraphics[width=\columnwidth]{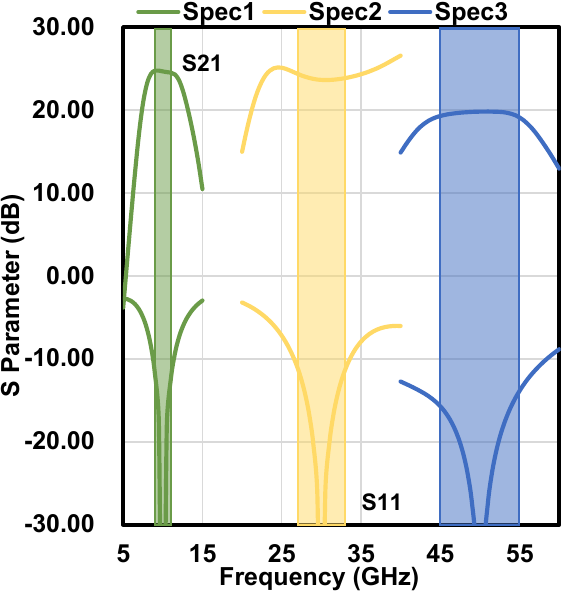}
    \caption{}
    \label{fig7:sub1}
  \end{subfigure}
  \hspace{0.005\linewidth}%
  \begin{subfigure}[t]{0.3\columnwidth}
    \centering
    \includegraphics[width=\columnwidth]{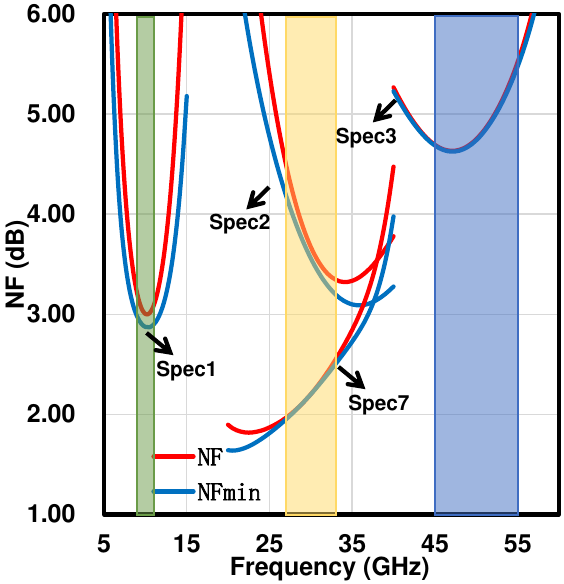}
    \caption{}
    \label{fig7:sub2}
  \end{subfigure}

  \medskip

  \begin{subfigure}[t]{0.3\columnwidth}
    \centering
    \includegraphics[width=\columnwidth]{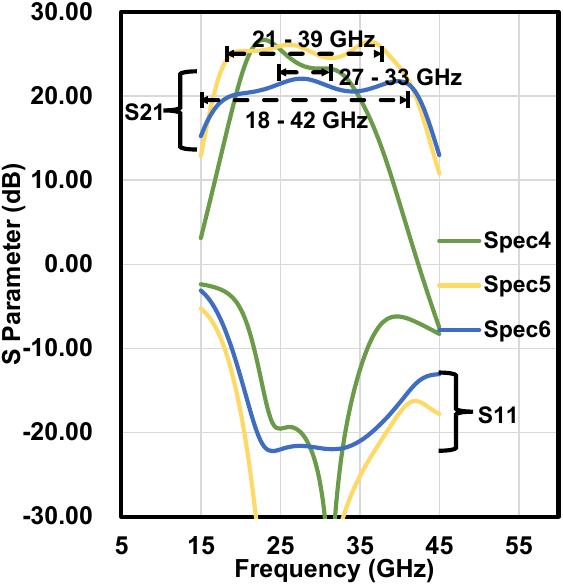}
    \caption{}
    \label{fig7:sub3}
  \end{subfigure}
  \hspace{0.005\linewidth}%
  \begin{subfigure}[t]{0.3\columnwidth}
    \centering
    \includegraphics[width=\columnwidth]{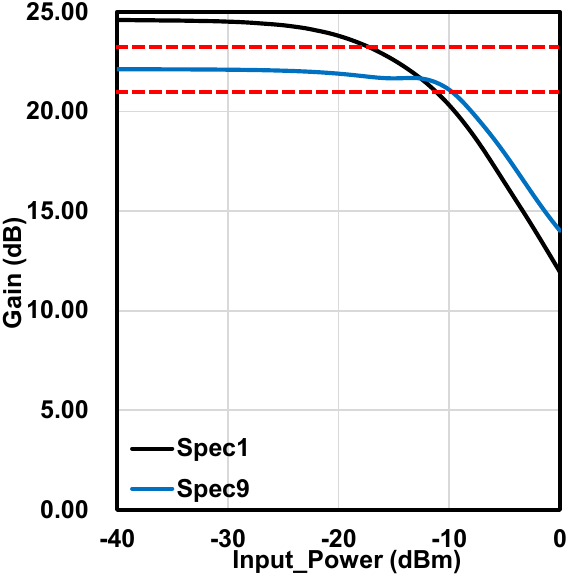}
    \caption{}
    \label{fig7:sub4}
  \end{subfigure}

  \caption{Design results of Specs.~1--9:
  (a) $\mathrm{S}_{11}$ and $\mathrm{S}_{21}$ of Specs.~1--3;
  (b) $\mathrm{NF}$ and $\mathrm{NF}_{\min}$ of Specs.~1--3 and~7;
  (c) $\mathrm{S}_{11}$ and $\mathrm{S}_{21}$ of Specs.~4--6;
  and (d) gain versus input power of Specs.~1 and~9.}
  \label{fig_example}
\end{figure}

\begin{figure}[!t]
  \centering
  \includegraphics[width=0.4\columnwidth]{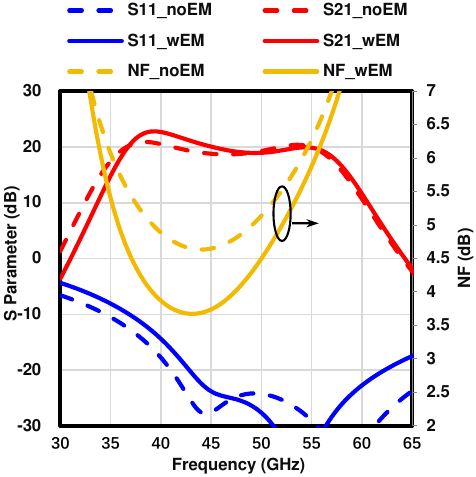}
  \caption{Post-EM results of Spec~3.}
  \label{fig_add1}
\end{figure}

\subsection{LNA and Driver Design Results}
To demonstrate the capability of the proposed RFAmpDesigner for wideband RF amplifier design, the results of the first eight tasks are summarized in Fig.~\ref{fig_example}. Fig.~\ref{fig7:sub1} and~\ref{fig7:sub2} present the S-parameters, $\mathrm{NF}$, and $\mathrm{NF}_{\mathrm{min}}$ for the first three tasks, confirming strong frequency scalability from 10 to 50~GHz. Fig.~\ref{fig7:sub2} further reports the $\mathrm{NF}$ and $\mathrm{NF}_{\mathrm{min}}$ of Spec~7 under a stringent low-$\mathrm{NF}$ requirement. Fig.~\ref{fig7:sub3} illustrates the $\mathrm{S}_{11}$ and $\mathrm{S}_{21}$ responses of Specs~4–6, showing that the framework accommodates user-specified bandwidths from $10\%$ to $80\%$. Finally, Fig.~\ref{fig7:sub4} depicts the power gain curve of Spec~1 and 9. Table~\ref{tab:params} lists the optimized component values for Spec~5. Fig.~\ref{fig_add1} depicts the post-EM results of Spec~3.

\subsection{Performance and Ablation Study}
\subsubsection{Effectiveness of the inner compiler}
Table~\ref{tab4} summarizes the results of all methods on easy tasks. For the first eight tasks, the active stages are fixed to control task difficulty and runtime by setting the power distribution to a balanced ratio of [0.4, 0.3, 0.3]. Vanilla BO and GA search the resulting 18-dimensional Fix Active space. To illustrate the hierarchical nature of RF design, BO and GA are also equipped with the stage 2 input matching network synthesis tool, reducing the search to a 16-dimensional space (Stage 2 Reduction). The two LLM baselines also operate in this 16-dimensional space. To simulate a realistic tight simulation budget, the time limit for easy tasks is empirically set to one hour. With each full evaluation taking approximately 20 seconds, this corresponds to a budget of 180 evaluations.

\begin{table}[!t]
\centering
\footnotesize
\caption{Experiment results of the first 8 tasks}
\label{tab4}
\setlength{\tabcolsep}{3pt}
\resizebox{\linewidth}{!}{%
\begin{tabular}{|c|c|c|c|c|c|c|c|c|c|c|c|c|c|c|}
\hline
\textbf{Spec} & \multicolumn{2}{c|}{\makecell{\textbf{GA-vanilla}\\(18-D)}} & \multicolumn{2}{c|}{\makecell{\textbf{BO-vanilla}\\(18-D)}} & \multicolumn{2}{c|}{\makecell{\textbf{GA-subtool}\\(16-D)}} & \multicolumn{2}{c|}{\makecell{\textbf{BO-subtool}\\(16-D)}} & \multicolumn{2}{c|}{\makecell{\textbf{AnalogCoder}\\(16-D)}} & \multicolumn{2}{c|}{\makecell{\textbf{ADO-LLM}\\(16-D)}} & \multicolumn{2}{c|}{\makecell{\textbf{Our Compiler}\\(8-D)}} \\
\cline{2-15}
 & \makecell{pass\\@1} & \makecell{Avg\\Time(s)\\(\# Sim)} & \makecell{pass\\@1} & \makecell{Avg\\Time(s)\\(\# Sim)} & \makecell{pass\\@1} & \makecell{Avg\\Time(s)\\(\# Sim)} & \makecell{pass\\@1} & \makecell{Avg\\Time(s)\\(\# Sim)} & \makecell{pass\\@1} & \makecell{Avg\\Time(s)\\(\# Sim)} & \makecell{pass\\@1} & \makecell{Avg\\Time(s)\\(\# Sim)} & \makecell{pass\\@1} & \makecell{Avg\\Time(s)\\(\# Sim)} \\
\hline
$S_1$ & 16/20 & \makecell{2408\\(121)} & 18/20 & \makecell{1777\\(86)} & 20/20 & \makecell{412\\(13)} & 20/20 & \makecell{547\\(17)} & 14/20 & \makecell{\textbf{240}\\(4)} & 20/20 & \makecell{571\\(14)} & \textbf{20/20} & \makecell{715\\(6)} \\
\hline
$S_2$ & 10/20 & \makecell{3096\\(155)} & 3/20 & \makecell{1146\\(55)} & 19/20 & \makecell{1531\\(49)} & 19/20 & \makecell{1016\\(31)} & 10/20 & \makecell{\textbf{363}\\(6)} & 18/20 & \makecell{1089\\(27)} & \textbf{20/20} & \makecell{744\\(4)} \\
\hline
$S_3$ & 0/20 & - & 0/20 & - & 0/20 & - & 1/20 & \makecell{3299\\(100)} & 0/20 & - & 0/20 & - & \textbf{5/20} & \makecell{\textbf{1755}\\(4)} \\
\hline
$S_4$ & 13/20 & \makecell{2541\\(127)} & 11/20 & \makecell{1621\\(78)} & 20/20 & \makecell{690\\(22)} & 20/20 & \makecell{619\\(19)} & 10/20 & \makecell{678\\(11)} & 17/20 & \makecell{\textbf{481}\\(12)} & \textbf{20/20} & \makecell{653\\(4)} \\
\hline
$S_5$ & 0/20 & - & 0/20 & - & 0/20 & - & 0/20 & - & 0/20 & - & 1/20 & \makecell{2455\\(61)} & \textbf{11/20} & \makecell{\textbf{1127}\\(17)} \\
\hline
$S_6$ & 0/20 & - & 0/20 & - & 0/20 & - & 0/20 & - & 0/20 & - & 0/20 & - & \textbf{5/20} & \makecell{\textbf{1586}\\(24)} \\
\hline
$S_7$ & 4/20 & \makecell{2944\\(147)} & 0/20 & - & 16/20 & \makecell{2182\\(69)} & 16/20 & \makecell{1360\\(41)} & 2/20 & \makecell{\textbf{502}\\(7)} & 15/20 & \makecell{1075\\(27)} & \textbf{18/20} & \makecell{1854\\(3)} \\
\hline
$S_8$ & 2/20 & \makecell{2713\\(138)} & 1/20 & \makecell{2401\\(116)} & 14/20 & \makecell{2639\\(85)} & 10/20 & \makecell{1814\\(56)} & 0/20 & - & 3/20 & \makecell{1544\\(39)} & \textbf{16/20} & \makecell{\textbf{662}\\(8)} \\
\hline
\end{tabular}
}%
\vspace{2pt}
\parbox{\linewidth}{%
    \footnotesize
    $^{*}$ - indicates that the baseline failed to complete
    the task within the time limit.\\
    $^{\dagger}$ \# Sim means the average number of
    full-chain simulation.
}
\end{table}

Examining the results vertically across three task clusters of increasing difficulty (Specs.~1--3, 4--6, and 7--8), the feasible region becomes progressively sparser as the center frequency increases, bandwidth widens, and power/noise constraints tighten. Accordingly, all methods exhibit degradation in both success rate and convergence time. Comparing horizontally, reducing the optimization space from 18 dimensions to 16 dimensions and finally to the compiler-assisted 8-dimensional resource space consistently improves both convergence efficiency and success rate. These results indicate that, under a limited simulation budget, sequential subproblem solving is substantially more effective than flat optimization.

The Stage2-assisted GA/BO baselines are designed as partial-tool ablations to isolate the contribution of inner-compiler decomposition. Since many failures in the easy regime originate from the critical input matching network, these baselines allow GA/BO to invoke only the Stage-2 sub-tool while retaining optimizer-based search for the remaining parameters. The improved performance on Specs.~4, 7, and 8 demonstrates that even partial decomposition can enhance conventional optimizers. The complete compiler is not combined with GA/BO in this comparison because, once the resource allocation is provided, it can directly generate feasible designs for many easy specifications with a single execution. In such cases, attaching GA/BO would mainly evaluate the compiler itself rather than the optimizer. Therefore, these baselines are intended to analyze the value of subproblem decomposition rather than replace the complete framework.

Turning to the LLM-based methods, AnalogCoder is highly efficient on narrowband tasks (Specs.~1, 2, and 4), completing them in at most one-third the time of other methods. This suggests that simple LLM reasoning about frequency requirements can provide useful initial design points. ADO-LLM generally requires longer completion times than AnalogCoder but achieves higher success rates, confirming that LLM reasoning alone is insufficient for reliable circuit optimization but can effectively assist optimization when integrated with appropriate tools. The compiler achieves the highest success rate across all tasks, indicating that for easy specifications, physics-driven subproblem decomposition and solving order remain the dominant factors.
To examine the dependence on the inner solver, Appendix~\ref{app:c} compares PSO/SA with deterministic solvers~(SLSQP and COBYQA) under the same fixed-current task setting. The alternatives show shorter reported average times on several tasks, but neither consistently matches the success rate of PSO/SA across $S_1$–$S_8$.

\begin{figure}[!t]
\centering
\includegraphics[width=0.45\columnwidth]{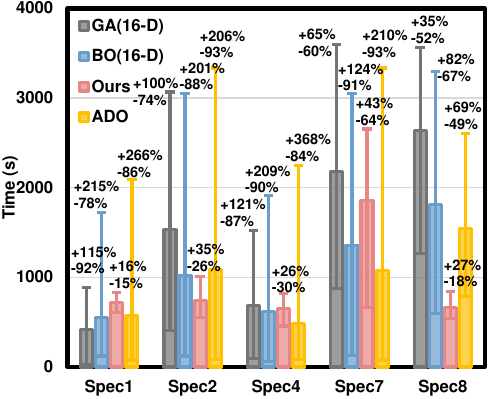}
\caption{Mean completion time with max-min variation range (\%) across tasks.}
\label{fig8}
\end{figure}

Fig.~\ref{fig8} shows the variance of all optimization methods on selected tasks. Since the core challenge in easy tasks is band planning, the distance from the initial point to the target operating band is critical. BO, GA, and ADO-LLM initialize randomly at the netlist level, where the probability that the superimposed MCR responses happen to fall near the target band is low, leading to high variance across seeds. By decomposing the problem and reducing dimensionality, the compiler's sensitivity to seed choice is substantially mitigated. The number of simulations required for convergence is listed below the average time. Notably, full-chain simulation in the compiler is invoked only after all subproblems have been successfully solved, so simulation count and wall-clock time are not directly proportional.

\begin{table}[!t]
\centering
\footnotesize
\caption{Experiment results of the last 2 tasks}
\label{tab5}
\setlength{\tabcolsep}{3pt}
\resizebox{\linewidth}{!}{%
\begin{tabular}{|c|c|c|c|c|c|c|c|c|}
\hline
\textbf{Methods} & \multicolumn{4}{c|}{$S_9$} & \multicolumn{4}{c|}{$S_{10}$} \\
\cline{2-9}
 & \makecell{pass\\@1} & \makecell{Avg Time\\(s)} & \makecell{Avg Prompt\\Tokens} & \makecell{Avg Completion\\Tokens} & \makecell{pass\\@1} & \makecell{Avg Time\\(s)} & \makecell{Avg Prompt\\Tokens} & \makecell{Avg Completion\\Tokens} \\
\hline
\hline
\multicolumn{9}{|c|}{\textbf{Main Experiment}} \\
\hline
\textbf{GA-Fixed-$K=1$} & 5/20 & 11728 & - & - & 1/20 & 15177 & - & - \\
\hline
\textbf{BO-Fixed-$K=1$} & 5/20 & 11424 & - & - & 2/20 & 8006 & - & - \\
\hline
\textbf{BO-Fixed-$K=3$} & 3/20 & 5915 & - & - & 6/20 & 9933 & - & - \\
\hline
\textbf{BO-Fixed-$K=5$} & 3/20 & 7280 & - & - & 6/20 & 6183 & - & - \\
\hline
\textbf{BO-Fixed-$K=10$} & 7/20 & 8879 & - & - & 9/20 & 11832 & - & - \\
\hline
\textbf{BO-Adaptive} & 16/20 & 6840 & - & - & 6/20 & 5394 & - & - \\
\hline
\makecell{\textbf{ADO-LLM-Fixed-$K=10$}\\\textbf{(Claude-Sonnet-4.5)}} & 16/20 & 4578 & 1812 & 1290 & 4/20 & 7985 & 1812 & 1504 \\
\hline
\makecell{\textbf{Ours-Claude-Sonnet-4.5}\\\textbf{Fixed-$K=10$}} & 19/20 & 6104 & 6059 & 2011 & 8/20 & 9090 & 12270 & 3347 \\
\hline
\textbf{Ours-Claude-Sonnet-4.5} & 19/20 & 5702 & 25759 & 3986 & 17/20 & 7428 & 40136 & 5398 \\
\hline
\hline
\multicolumn{9}{|c|}{\textbf{Cross-Backbone Evaluation}} \\
\hline
\textbf{Ours-Gemini-2.5-Flash} & 5/5 & 4521 & 36215 & 5005 & 3/5 & 2014 & 18840 & 2609 \\
\hline
\textbf{Ours-DeepSeek-V3.2} & 5/5 & 6114 & 62388 & 5652 & 5/5 & 6788 & 71939 & 6576 \\
\hline
\textbf{Ours-GPT-5.1} & 5/5 & 5611 & 62069 & 4963 & 3/5 & 13064 & 126893 & 9106 \\
\hline
\textbf{Ours-Qwen3-480B-A35B} & 4/5 & 5624 & 47623 & 2254 & 3/5 & 8302 & 29488 & 2344 \\
\hline
\textbf{Ours-Qwen3-30B-A3B} & 2/5 & 3926 & 65454 & 2953 & 1/5 & 2670 & 7454 & 1805 \\
\hline
\hline
\multicolumn{9}{|c|}{\textbf{Ablation Study (backbone: Claude-Sonnet-4.5)}} \\
\hline
\makecell{\textbf{Multi-Agent w/ EB}} & 5/5 & 6100 & 8643 & 2399 & 5/5 & 6469 & 81884 & 7828 \\
\hline
\makecell{\textbf{Multi-Agent}} & 5/5 & 4785 & 25759 & 3986 & 5/5 & 6577 & 40136 & 5398 \\
\hline
\makecell{\textbf{Single Agent}} & 5/5 & 6024 & 161150 & 6816 & 5/5 & 8308 & 55785 & 3658 \\
\hline
\makecell{\textbf{w/o Refiner}} & 4/5 & 8681 & 8200 & 2719 & 2/5 & 16015 & 15135 & 3672 \\
\hline
\makecell{\textbf{w/o ReAct}} & 5/5 & 10411 & 92489 & 3344 & 3/5 & 13816 & 96341 & 3364 \\
\hline
\end{tabular}
}%
\end{table}

\begin{figure}[!t]
  \centering
  \includegraphics[width=0.45\columnwidth]{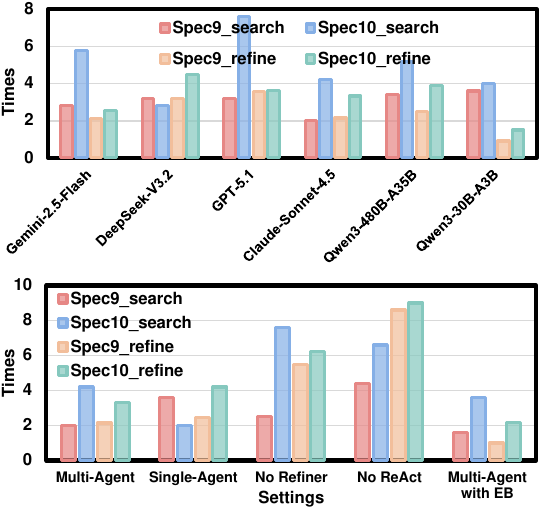}
  \caption{Mean number of search and refine calls per seed of different models (top); Mean number of search and refine calls per seed of different configuration of RFAmpDesigner (down).}
  \label{fig11}
\end{figure}

\subsubsection{Validity of LLM-guided tool orchestration}
Table~\ref{tab5} reports the results on Specs~9--10. For hard-regime tasks, the key challenge is not only finding feasible resource allocations but also allocating the simulation budget between the expensive Search tool and the low-cost Refine tool. Therefore, different Search-Refine scheduling strategies are compared under a five-hour budget.
As shown in Table~\ref{tab5}, $K=10$ gives the highest success rate on both specifications among the fixed-$K$ BO baselines. BO-Adaptive increases $S_9$ success from 7/20 to 16/20, but reduces $S_{10}$ success from 9/20 to 6/20. The effect of the tested cost-based scheduling rule differs between the two tasks, indicating that manually selected ratios cannot adapt to varying design difficulties. RFAmpDesigner obtains 19/20 and 17/20, while its fixed-$K$ version obtains 19/20 and 8/20, respectively. Adaptive scheduling thus improves the observed success rate on $S_{10}$, with no change on $S_9$. Since all methods share the same compiler and resource space, these results support the benefit of the LLM controller over the tested optimizers and non-LLM adaptive scheduler under the same time limit.

Fig.~\ref{fig11} further compares numbers of Search and Refine calls. The search iterations are broadly similar across LLM-based methods and increase with task difficulty. Overall, Claude-Sonnet-4.5 offers the best balance among success rate, runtime, and token cost.

Without the compiler, the original AnalogCoder and ADO-LLM are almost unable to complete these tasks. 
This motivate the resource-level interface used in this study.
The open-source Qwen models show a clear performance gap. Qwen3-480B-A35B approaches frontier closed-source models, while the smaller Qwen3-30B-A3B suffers from tool-call formatting and RF-reasoning errors. This indicates that both model capability and tool-use specialization are important for reliable RF-oriented workflows.

\newsavebox{\mytablebox}

\begin{table}[!t]
\centering
\footnotesize

\caption{Comparison of effective latency and monetary cost across different backbones}
\label{tab6}

\begin{lrbox}{\mytablebox}
\begin{tabular}{|c|c|c|c|c|}
\hline
\textbf{Model} & \multicolumn{2}{c|}{$S_9$} & \multicolumn{2}{c|}{$S_{10}$} \\
\cline{2-5}
 & \makecell{Eff Latency(s)} & \makecell{Eff Cost (USD)} & \makecell{Eff Latency(s)} & \makecell{Eff Cost (USD)}\\
\hline
\textbf{Gemini-2.5-Flash} & 4521 & 0.023 & 3356 & 0.020 \\
\hline
\textbf{DeepSeek-V3.2} & 6114 & 0.018 & 6788 & 0.020 \\
\hline
\textbf{GPT-5.1} & 5611 & 0.127 & 21773 & 0.416 \\
\hline
\textbf{Claude-Sonnet-4.5} & 6002 & 0.137 & 8739 & 0.201 \\
\hline
\textbf{Qwen3-480B-A35B} & 7030 & 0.061 & 13837 & 0.056 \\
\hline
\textbf{Qwen3-30B-A3B} & 9815 & 0.021 & 13352 & 0.008 \\
\hline
\end{tabular}
\end{lrbox}

\usebox{\mytablebox}

\vspace{2pt}

\parbox{\wd\mytablebox}{%
    \raggedright
    $^{*}$ Effective latency and cost are calculated
    by dividing the pass@1 rate.
}
\end{table}

Table~\ref{tab6} reports the effective time and LLM cost. The LLM overhead amounts to a fraction of a dollar per task. Under a limited simulation time budget, this cost is more economical than provisioning additional machines for brute-force search.

\subsubsection{Ablation of Multi-Agent Framework}
Finally, to substantiate the efficiency benefit of the multi-agent decomposition, ablation studies are conducted using Claude-Sonnet-4.5 as the backbone. 

Ablation 1: Multi-Agent vs. Single-Agent. The Single-Agent variant achieves a comparable success rate, but consumes far more prompt tokens. For example, its prompt token consumption on Spec~9 (161k) is more than 6 times higher than our proposed framework (25k). This indicates that multi-agent structure substantially reduces reasoning burden and improves token efficiency by agentic decomposition.

Ablation 2: Effect of the Refiner. Removing the Refiner loop causes a pronounced degradation, especially on Spec~10 where success drops from 100\% to 40\%. Although token usage decreases (as the Refiner is the main token contributor), search becomes less effective. Fig.~\ref{fig11} shows the number of refinement nearly doubles. This suggests that, without expert-guided tool orchestration, the workflow relies more heavily on repeated candidate search and becomes less efficient.

Ablation 3: Reasoning Mode. The “without ReAct” variant exhibits a higher observed average runtime and a lower success rate on Spec~10. This result supports the usefulness of interleaved reasoning and action for interpreting simulation feedback and revising the search strategy.

Ablation 4: Role of the EB. While the inclusion of EB yields only minor changes in overall success rate and wall-clock time on these tasks, it reduced the number of refinement calls in the evaluated runs. Compared with the corresponding multi-agent setting without EB, EB reduced the mean number of refinement calls by at least 40\%. Retrieved historical design traces provide an effective pruning signal, enabling fewer redundant trials.

\begin{figure}[!t]
  \centering
  \includegraphics[width=0.5\columnwidth]{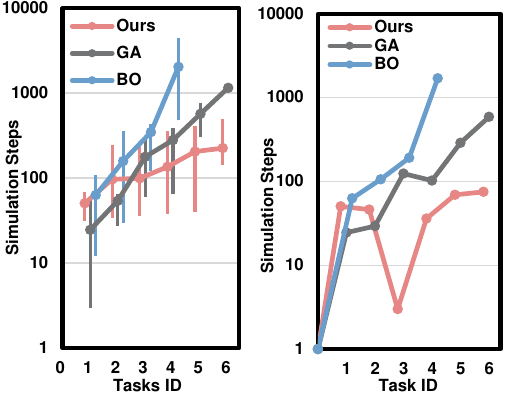}
  \caption{Cumulative simulation cost with error bar over 5 seeds (left); Average per-task simulation cost over 5 seeds (right).}
  \label{fig12}
\end{figure}

\begin{figure}[!t]
  \centering
  \includegraphics[width=0.4\columnwidth]{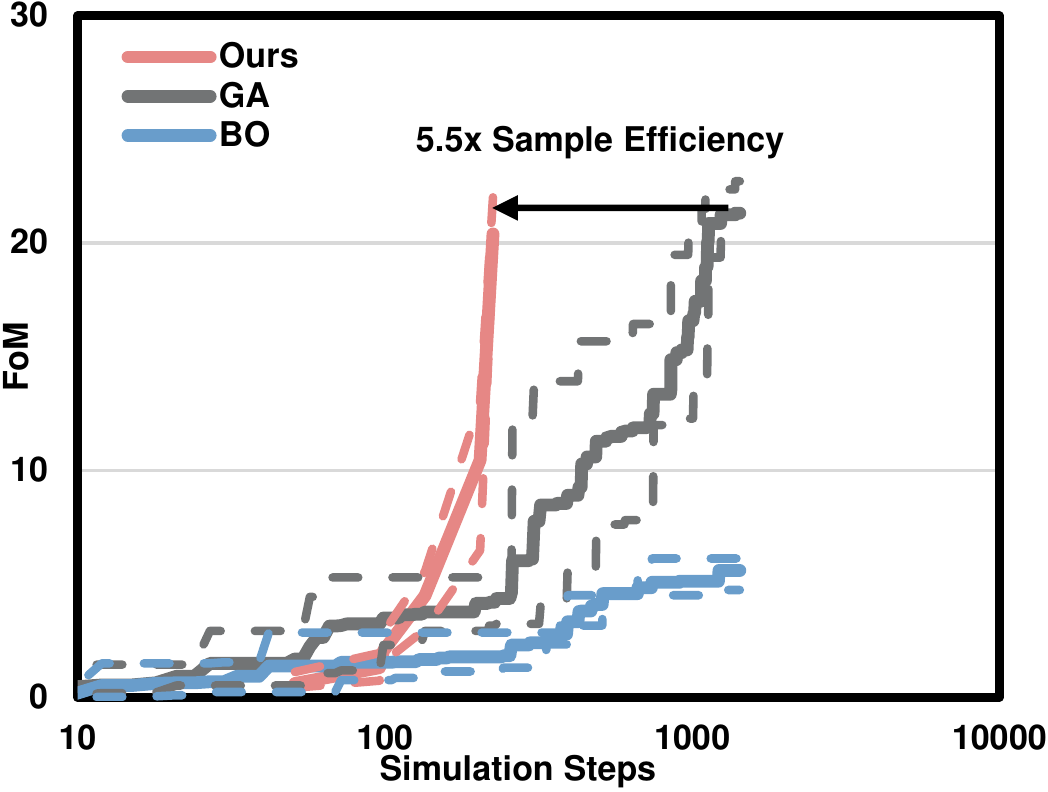}
  \caption{FoM Evolution vs. Simulation Cost.}
  \label{fig13}
\end{figure}

\subsection{Comparison of Figure-of-Merit Trade-offs and Exploitation Strategies}
To examine how different methods exploit previously explored points when pushing toward extreme performance requirements after satisfying the band constraint, a sequential task series is constructed based on Spec~4 with progressively increasing gain requirements (from 20 to 45 dB) and warm-start setting, using the same figure-of-merit (FoM) definition as in~\cite{LNAFoM}. The left panel of Fig.~\ref{fig12} shows that RFAmpDesigner avoids the exponential cost growth observed in vanilla GA and BO (clearly visible on the logarithmic scale), and grows substantially more slowly than GA and BO over this task sequence. The underlying mechanism is revealed by the “sawtooth” pattern in the right panel of Fig.~\ref{fig12}. As task difficulty increases, RFAmpDesigner explicitly exploits previously explored points, producing sharp cost reductions (the valleys enabled by memory reuse). Importantly, this acceleration does not sacrifice design quality. Fig.~\ref{fig13} shows the evolution of FoM across tasks. RFAmpDesigner reaches comparable FoM values with a 5.5$\times$ improvement in sampling efficiency.

\begin{figure}[!t]
  \centering
  \includegraphics[width=0.5\columnwidth]{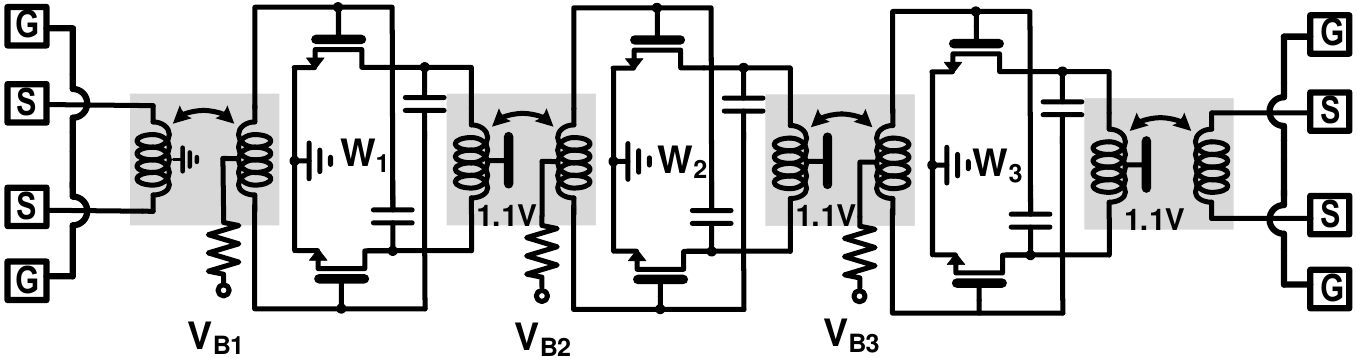}
  \caption{PA schematic}
  \label{fig14}
\end{figure}

\begin{figure}[!t]
  \centering
  \subfloat[]{%
  \includegraphics[width=0.7\columnwidth]{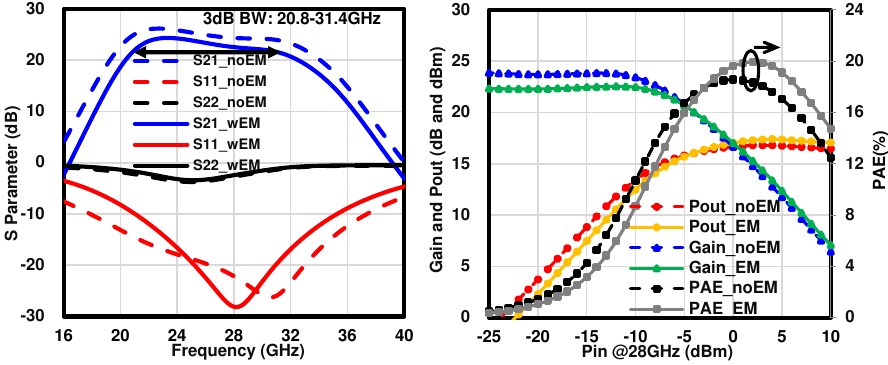}
  \label{fig15:sub1}
}
\hfill


\subfloat[]{%
  \includegraphics[width=0.4\columnwidth]{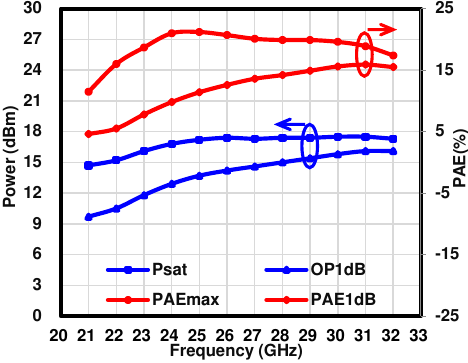}
  \label{fig15:sub2}
}
\hfill

\caption{Design results of PA before and after EM simulation: a) $\mathrm{S_{21}}$, $\mathrm{S_{11}}$ and $\mathrm{S_{22}}$; Pin vs Pout, PAE and Power Gain; b) Post-EM broadband performance.}
\label{fig15}
\end{figure}

\newsavebox{\PATableBox}

\begin{table}[!t]
\centering
\footnotesize
\caption{Comparison of CMOS differential PA}
\label{tablePA}

\begin{lrbox}{\PATableBox}
\begin{tabular}{|c|c|c|c|}
\hline
Circuit & \multicolumn{3}{c|}{PA} \\
\cline{2-4}
 & This work & \cite{competitor_B} & \cite{PAcite2} \\
\hline
\makecell{CMOS} & 65nm & 40nm & 65nm \\
\hline
VDD (V) & 1.1 & 1.2 & 1.2 \\
\hline
Gain (dB) & 22.5 & 26.3 & 23.3 \\
\hline
Freq (GHz) & 28 & 28 & 28 \\
\hline
$BW_{-3dB}$ (GHz) & 10.6 & 15 & 9 \\
\hline
$P_{sat}$ (dBm) & 17.4 & 17 & 17 \\
\hline
OP1dB (dBm) & 15 & 9 & 8.3 \\
\hline
$PAE_{1dB}$ (\%) & 14.2 & 22.5 & 15 \\
\hline
Time (h) & 2.5 & 22 & $48^{\dagger}$ \\
\hline
\end{tabular}
\end{lrbox}

\usebox{\PATableBox}

\vspace{2pt}

\parbox{\wd\PATableBox}{%
    \raggedright
    $^{\dagger}$ 200 designs were obtained from multi-object optimization.
}

\end{table}

\begin{figure}[!t]
  \centering

  \begin{subfigure}{\columnwidth}
    \centering
    \includegraphics[width=0.5\columnwidth]
    {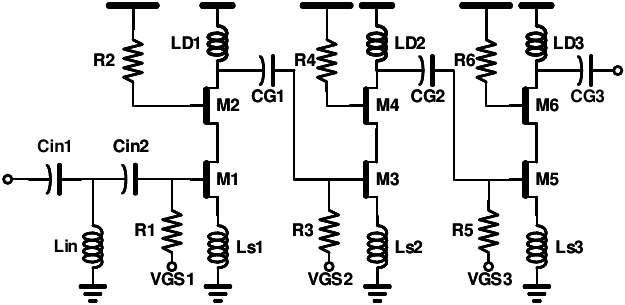}
    \caption{Schematic of single-ended LNA.}
    \label{figlnacasestudy1}
  \end{subfigure}

  \vspace{3mm}

  \begin{subfigure}{\columnwidth}
    \centering
    \includegraphics[width=0.4\columnwidth]
    {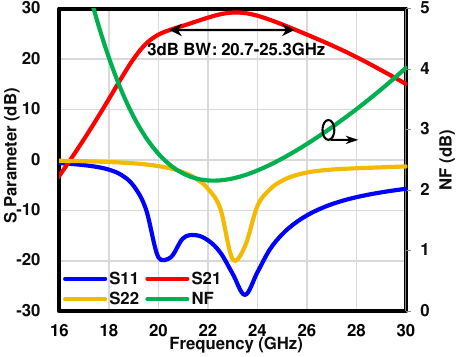}
    \caption{Post-EM S-parameters and NF.}
    \label{figlnacasestudy2}
  \end{subfigure}

  \caption{Single-ended LNA case study.}
  \label{fig:single_lna_case_study}
\end{figure}

\subsection{Extended Results and Discussion about EM and Generality}
A three-stage differential CMOS PA is designed using RFAmpDesigner (shown in Fig.~\ref{fig14}), and the design results are presented in Fig.~\ref{fig15}. The EM simulation results are obtained by applying the bisection method from~\cite{hexy} to each MCR after sizing optimization. 
The sizing process takes 109 minutes. The EM optimization of the transformers requires 43 iterations in total, with each iteration taking about 1 minute. Therefore, the total runtime is 152 minutes, which is about 2.5 hours. A comparison with prior works is summarized in Table~\ref{tablePA}.

To illustrate how the compiler-based decomposition approach can be adapted beyond the primary target family, a single-ended LNA case study shown in Fig.~\ref{figlnacasestudy1} is also provided. The LNA topology is similar to those in~\cite{LNAcite1,LNAcite2}, where the interstage matching network is implemented using an LC network instead of an MCR. A key difference is that all active stages in this design use a cascode structure. Following the idea of the Compiler, the design problem is decomposed into a sequence of subproblems and solved sequentially. The same sub-optimization algorithm and hyperparameters are used as before. The inductors are automatically EM-optimized using PCell and EMX scripts together with the bisection method described above. This process requires 34 evaluations in total, with each evaluation taking about 1 minute. The final errors of all inductors are within 2.5\%. The post-EM results are shown in Fig.~\ref{figlnacasestudy2}. The sizing and EM optimization of inductors takes 0.75 hour in total.

To better align schematic and layout results, active-device parasitics are extracted in advance, the finite Q of passive devices is included in schematic simulation, and resistors and capacitors are implemented with PDK devices. The passive Q is set to 10 in this process, and the mutual inductance of each transformer is limited to $M \leq min(L_1, L_2)$. The MCR model is treated as reliable only when the operating frequency is well below its estimated self-resonant frequency. This condition becomes difficult to satisfy in high-frequency designs with large inductance and strong coupling. For Spec3, one solution uses $k = 0.7$ and $L = 500$ pH, which leads to a self-resonant frequency of about 64 GHz and severe post-EM degradation, so the design fails to meet the target specifications. Another solution with moderate $k$ and $L$ shows only small post-EM changes.
These measures provide only open-loop physical awareness. As Spec3 shows, a realizable geometry does not guarantee a valid lumped model, and EM results are not fed back into the search. A component-level closed loop would require two checks: a process-specific EM surrogate of the MCR to check physical realizability, and an electrical feasibility check in the inner compiler based on the EM-consistent behavior. Unrealizable requirements should be attributed to the outer-level allocation via the diagnostic feedback path. Even so, sign-off still requires full-layout EM simulation, where inter-stage coupling, interconnects, and return paths may invalidate verified component-level results. Resolving such deviations calls for a new interaction and intervention mechanism across the entire loop, which is beyond the scope of this work. Since both extensions increase the per-candidate cost and shrink the feasible region, efficient sizing remains their prerequisite.

The proposed method is built on a stage-wise interface, where the LLM allocates resource budgets and the compiler tunes circuit parameters through stage-specific tools. For multi-stage circuits, adding stages mainly increases the number of budget variables, while the same decision loop can be reused. 

For the topology adaptations considered in this work, the outer decision loop and sub-optimization script templates can be reused. Within the same CMOS process, the device lookup table is also reusable when the required device configurations and operating conditions are covered by the existing characterization. Beyond updating the full-circuit evaluation setup, the main compiler-specific tasks are extracting the critical-stage DUT netlists from the full-circuit netlist and reconfiguring the optimization-variable lists. The Stage 2 testbench and OCEAN templates can be retained. The Stage 3 MCR-based multi-fidelity functions are reusable within the target amplifier family. These adaptations still require manual configuration and verification, while fully automated topology adaptation remains future work.

\section{Conclusion}
\label{sec5}
This paper presents RFAmpDesigner, a two-level optimization framework for automated schematic-level parameter sizing of differential CMOS RF amplifiers. The method is demonstrated on low-noise, driver, and power amplifiers, covering the design flow from specifications to schematics.
The key idea is to decompose the high-dimensional optimization problem into two levels. At the outer level, an LLM makes design decisions in a physically meaningful coordinate system. At the inner level, domain knowledge is used to further decompose the problem into a sequence of sub-optimization tasks. 
Experiments support the effectiveness of compiler decomposition on $S_1$–$S_8$ under fixed active stages and an incremental benefit of the complete LLM controller over the tested baselines on the resource-allocation tasks $S_9$–$S_{10}$. Selected post-EM studies assess the transfer of schematic designs to EM-extracted passive models. Integrating physical-rule checks and EM feedback into the optimization loop remains an extension of the present framework.

\bibliographystyle{ACM-Reference-Format}
\bibliography{main}

\appendix
\section{Controlled Reinforcement-Learning Evaluation}
\label{app:rl}

\subsection{Environment and Evaluation Setting}

To evaluate reinforcement learning under a tractable yet representative setting, proximal policy optimization (PPO) is applied to the stage~3 bandwidth-planning subproblem. The active stages are fixed, while the three MCRs are optimized using the same device database and analytical models as those used in the main experiments. The overall frequency response is obtained by cascading the three MCR responses.

Following the RL formulation in~\cite{Autockt}, the design vector contains 21 passive variables, corresponding to seven parameters for each MCR. The variables are discretized with step sizes of 0.02 for the coupling coefficient, 10~pH for inductance, 20~$\Omega$ for resistance, and 10~fF for capacitance. Their ranges follow Table I, except that the upper bound of the inductance is extended to 2000~pH.

The PPO agent is trained on 50 randomly generated specifications, with all evaluation specifications excluded from training. The held-out test set contains the original specifications S1--S5 and a modified new specification S6, obtained by increasing the fractional bandwidth of S5 to 40\%. The dense training reward penalizes center-frequency and bandwidth errors, incomplete 3-dB bandwidth coverage, gain degradation, minimum-gain violation, and ripple violation. To decouple the evaluation from the reward design, a run is regarded as successful only when the same criteria used for GA and CMA-ES are satisfied, i.e., the optimization cost is below 100 and the required 3-dB bandwidth is fully covered.

PPO is implemented using Ray/RLlib with a rollout horizon of 500, a training batch size of 64\,000, a sample batch size of 500, an SGD minibatch size of 1000, 20 SGD iterations per update, and a learning rate of $10^{-4}$. The policy network contains two hidden layers of 64 units, and 128 rollout workers are used. Each checkpoint is evaluated using five random seeds with a maximum test horizon of 2000 environment steps.

\subsection{Results}
\begin{figure}[!t]
  \centering
  \includegraphics[width=0.8\columnwidth]{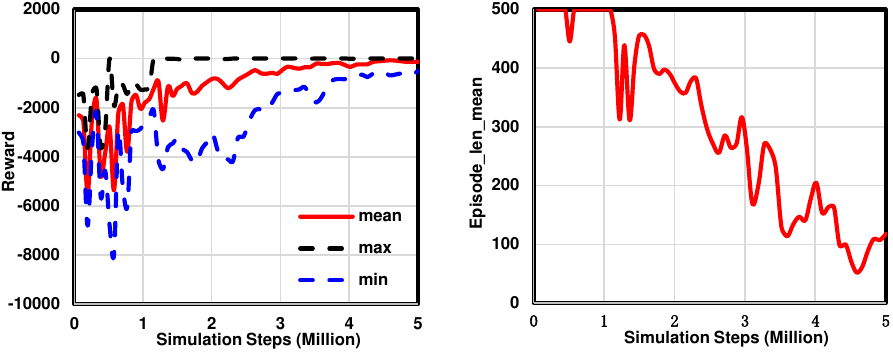}
\caption{(Left) Reward vs. steps; (Right) Mean episodic length vs. steps}
\label{fig:rl_trainin}
\end{figure}

Fig.~\ref{fig:rl_trainin} shows that the episodic reward progressively increases while the episode length decreases during training. Across the six specifications and five evaluation seeds, the numbers of successful runs at 1, 2, 3, 4, and 5 million training steps are 2/30, 15/30, 27/30, 30/30, and 30/30, respectively. Full test-set coverage is first obtained at 4 million steps, while further training generally reduces the required test-time rollout length. Detailed checkpoint results are summarized in Table~\ref{tab:rl_results}.

Although the trained PPO policy requires substantially fewer test-time environment steps than GA and CMA-ES. These results demonstrate that PPO can effectively solve the constrained bandwidth-planning subproblem when sufficient offline interactions are available. A full-chain RL evaluation using the same high-fidelity environment, however, was beyond the computational budget of the present study. Sample-efficient, surrogate-assisted, or transfer-based RL techniques may reduce this cost in future work.

\begin{table}[t]
\centering
\caption{Test results on Evaluation set of all methods.}
\label{tab:rl_results}
\scriptsize
\setlength{\tabcolsep}{6pt}
\renewcommand{\arraystretch}{1.05}
\setlength{\tabcolsep}{3pt}
\resizebox{\linewidth}{!}{%
\begin{tabular}{l*{6}{cc}c}
\hline
\textbf{Method}
& \multicolumn{2}{c}{\textbf{S1}}
& \multicolumn{2}{c}{\textbf{S2}}
& \multicolumn{2}{c}{\textbf{S3}}
& \multicolumn{2}{c}{\textbf{S4}}
& \multicolumn{2}{c}{\textbf{S5}}
& \multicolumn{2}{c}{\textbf{S6}}
& \textbf{Overall} \\
\cline{2-13}
& \textbf{Succ.} & \textbf{Steps}
& \textbf{Succ.} & \textbf{Steps}
& \textbf{Succ.} & \textbf{Steps}
& \textbf{Succ.} & \textbf{Steps}
& \textbf{Succ.} & \textbf{Steps}
& \textbf{Succ.} & \textbf{Steps}
& \textbf{Succ.} \\
\hline

CMA-ES
& 5/5 & 83
& 5/5 & 1516
& 5/5 & 4014
& 5/5 & 1225
& 5/5 & 4156
& 5/5 & 4237
& 30/30 \\

GA
& 5/5 & 205
& 5/5 & 1685
& 5/5 & 3182
& 5/5 & 1172
& 5/5 & 2591
& 5/5 & 2340
& 30/30 \\

PPO (1M)
& 0/5 & --
& 2/5 & 122
& 0/5 & --
& 0/5 & --
& 0/5 & --
& 0/5 & --
& 2/30 \\

PPO (2M)
& 0/5 & --
& 5/5 & 49
& 4/5 & 256
& 4/5 & 541
& 2/5 & 324
& 0/5 & --
& 15/30 \\

PPO (3M)
& 2/5 & 492
& 5/5 & 40
& 5/5 & 70
& 5/5 & 177
& 5/5 & 110
& 5/5 & 77
& 27/30 \\

PPO (4M)
& 5/5 & 135
& 5/5 & 13
& 5/5 & 22
& 5/5 & 69
& 5/5 & 84
& 5/5 & 30
& 30/30 \\

PPO (5M)
& 5/5 & 119
& 5/5 & 9
& 5/5 & 23
& 5/5 & 38
& 5/5 & 38
& 5/5 & 8
& 30/30 \\
\hline
\end{tabular}
}%

\vspace{1mm}
\begin{minipage}{0.98\textwidth}
\footnotesize
\textit{Note:} Succ. denotes the number of successful runs out of five. Steps denotes the mean number of test-time environment steps over successful runs. 
\end{minipage}
\end{table}

\section{Resource Vectors Distributions}
\begin{figure}[!t]
    \centering

    \begin{subfigure}[t]{0.48\textwidth}
        \centering
        \includegraphics[width=\linewidth]
            {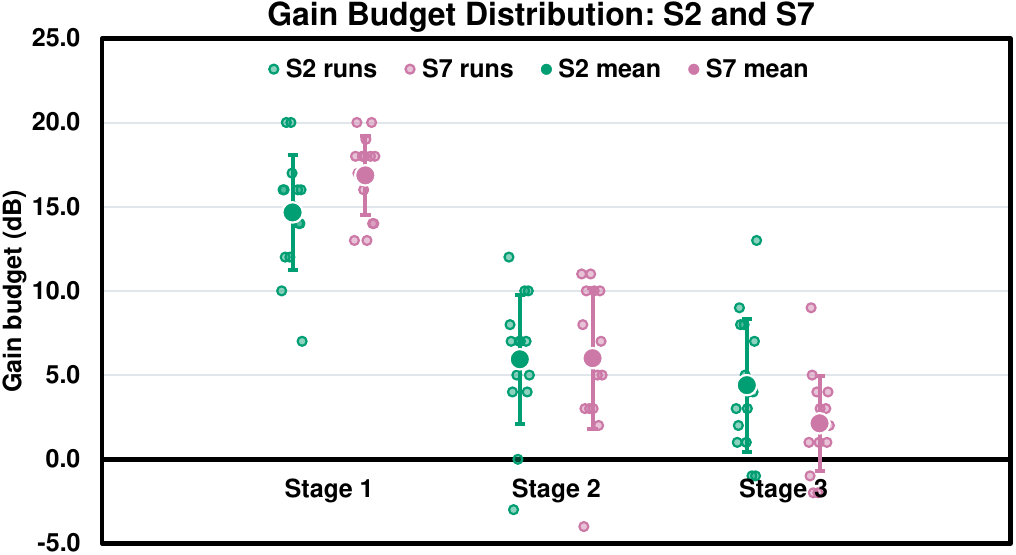}
        \caption{}
        \label{fig:gain-s2-s7}
    \end{subfigure}
    \hfill
    \begin{subfigure}[t]{0.48\textwidth}
        \centering
        \includegraphics[width=\linewidth]
            {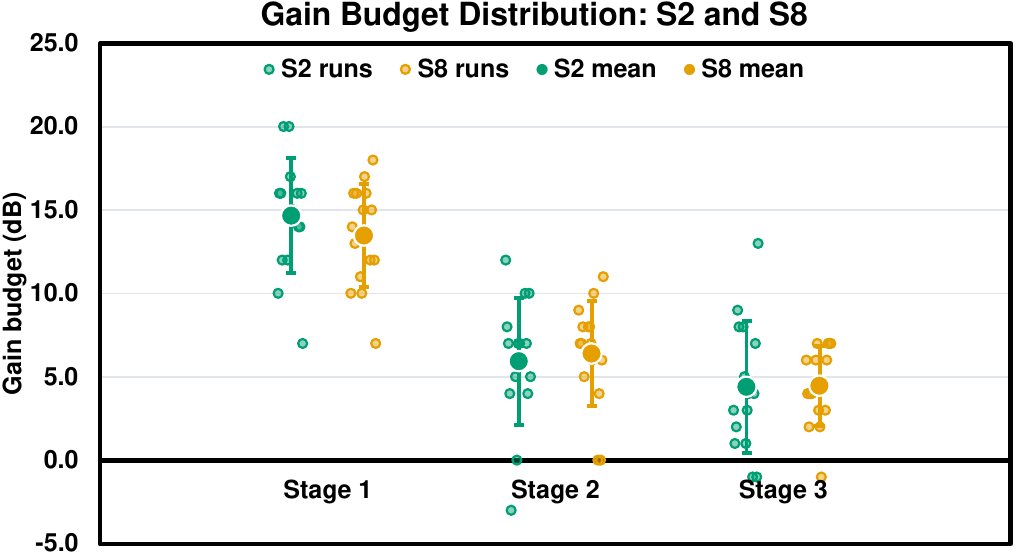}
        \caption{}
        \label{fig:gain-s2-s8}
    \end{subfigure}

    \medskip

    \begin{subfigure}[t]{0.48\textwidth}
        \centering
        \includegraphics[width=\linewidth]
            {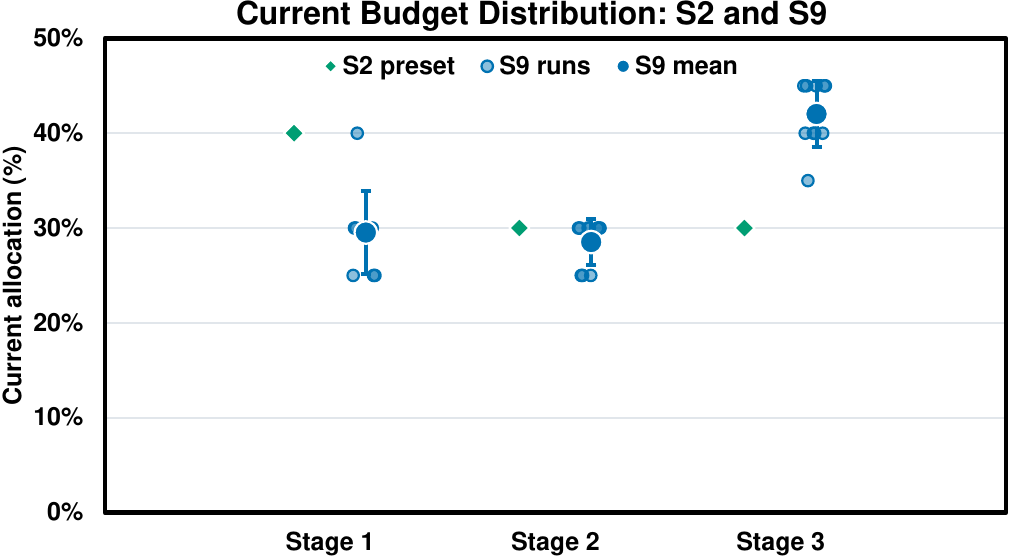}
        \caption{}
        \label{fig:current-s2-s9}
    \end{subfigure}
    \hfill
    \begin{subfigure}[t]{0.48\textwidth}
        \centering
        \includegraphics[width=\linewidth]
            {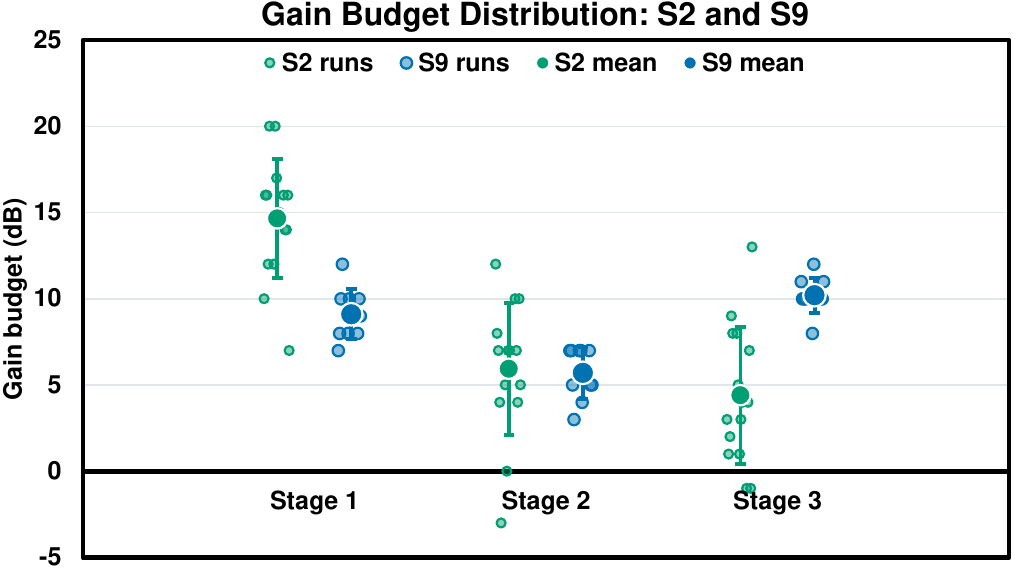}
        \caption{}
        \label{fig:gain-s2-s9}
    \end{subfigure}

    \medskip

    \begin{subfigure}[t]{0.48\textwidth}
        \centering
        \includegraphics[width=\linewidth]
            {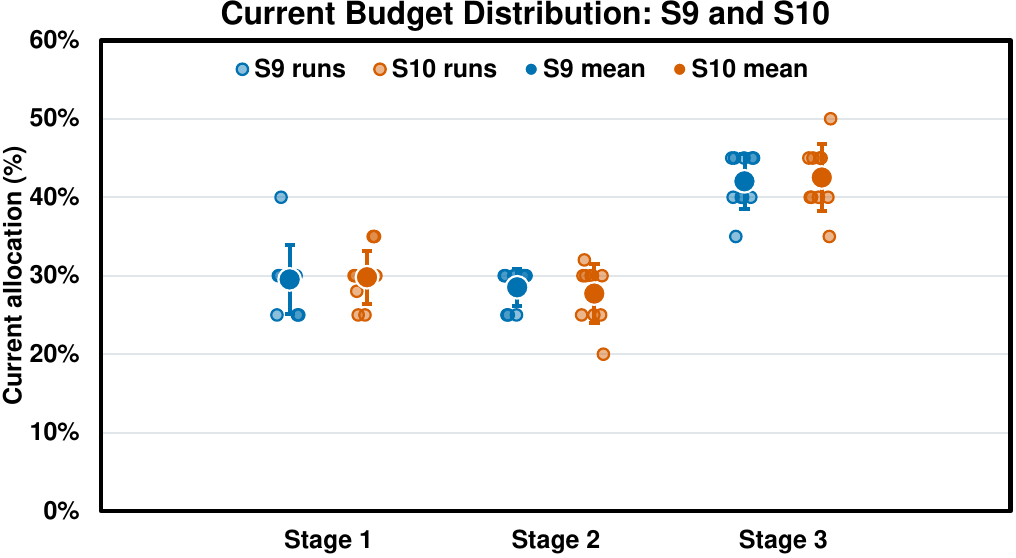}
        \caption{}
        \label{fig:current-s9-s10}
    \end{subfigure}
    \hfill
    \begin{subfigure}[t]{0.48\textwidth}
        \centering
        \includegraphics[width=\linewidth]
            {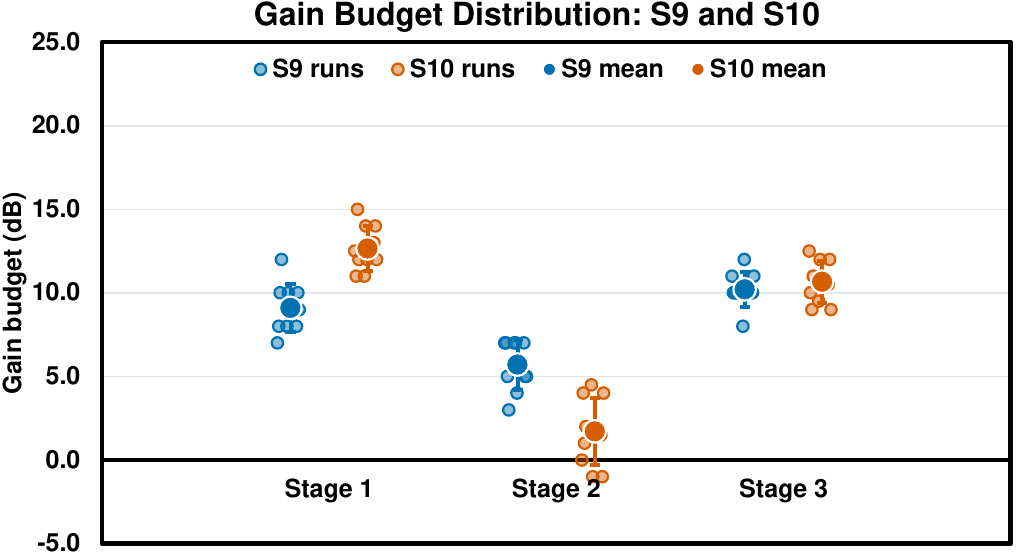}
        \caption{}
        \label{fig:gain-s9-s10}
    \end{subfigure}

    \caption{Current and gain allocations in the resource vectors.
    (a) Gain allocation for S2 and S7.
    (b) Gain allocation for S2 and S8.
    (c) Current allocation and (d) gain allocation for S2 and S9.
    (e) Current allocation and (f) gain allocation for S9 and S10.
    S2 current markers in (c) show the fixed allocation.
    For the other series, small markers show individual
    allocations and large markers show the means.
    Error bars show $\pm 1$ sample standard deviation.}
    \label{fig:resource-budgets}
\end{figure}

Fig.~\ref{fig:resource-budgets} shows the current allocation $\boldsymbol{\rho}_P$ and gain allocation $\boldsymbol{\rho}_G$ in the resource vectors $\mathbf{r}$ under different noise, current, and linearity constraints. The current allocation is fixed at $(40\%, 30\%, 30\%)$ in $S_2$, $S_7$, and $S_8$.

Fig.~\ref{fig:resource-budgets}(a) compares $S_2$ and $S_7$. The noise figure limit decreases from 5 to 2.5~dB. The mean first-stage gain allocation increases from 14.67 to 16.87~dB, while the mean third-stage gain allocation decreases from 4.40 to 2.13~dB. The second-stage gain allocation remains nearly unchanged. This shift is consistent with the use of higher first-stage gain to reduce the noise contribution of later stages.

Fig.~\ref{fig:resource-budgets}(b) compares $S_2$ and $S_8$. The total current budget decreases from 30 to 15~mA. The gain allocations remain broadly similar. The mean gain allocated to the first stage decreases from 14.67 to 13.47~dB, while slightly more gain is assigned to the second and third stages. Under the lower power budget, the allocation tends to reduce the gain required from the first stage and let the later stages provide more gain.

Figs.~\ref{fig:resource-budgets}(c) and~\ref{fig:resource-budgets}(d) compare $S_2$ and $S_9$. The minimum required $\mathrm{IP}_{1\mathrm{dB}}$ increases from $-25$ to $-15$~dBm. The other benchmark specifications remain unchanged. The mean current allocation in $S_9$ is $(29.5\%, 28.5\%, 42.0\%)$, with more current assigned to the final stage than in $S_2$. The mean gain allocated to the first stage decreases from 14.67 to 9.10~dB, while that allocated to the third stage increases from 4.40 to 10.20~dB. The second-stage gain allocation remains similar. This redistribution shifts both current and gain from the first stage to the third stage. With suitable near-threshold biasing, the reduced first-stage current can support gain expansion. The lower gain before the third stage also reduces its input drive, easing the linearity burden on that stage.

Figs.~\ref{fig:resource-budgets}(e) and~\ref{fig:resource-budgets}(f) compare $S_9$ and $S_{10}$. Both tasks have the same $\mathrm{IP}_{1\mathrm{dB}}$ requirement. The noise figure limit decreases from 5 to 3.5~dB. The current allocations remain similar, with a ratio of approximately $3:3:4$. Each of the first two stages receives less current than the final stage. With suitable near-threshold biasing, this allocation can support gain expansion in the first two stages. This gain expansion can partly compensate for gain compression in the final stage. The mean gain allocated to the first stage increases from 9.10 to 12.65~dB. The second-stage gain allocation decreases from 5.70 to 1.70~dB, while the third-stage gain allocation remains similar. This follows the same trend as the $S_2$ and $S_7$ comparison. In both cases, the tighter noise constraint favors more first-stage gain to suppress the noise contribution of later stages.

\section{Comparison of Inner Solvers}
\label{app:c}
\begin{figure}[!t]
  \centering
  \includegraphics[width=0.45\columnwidth]{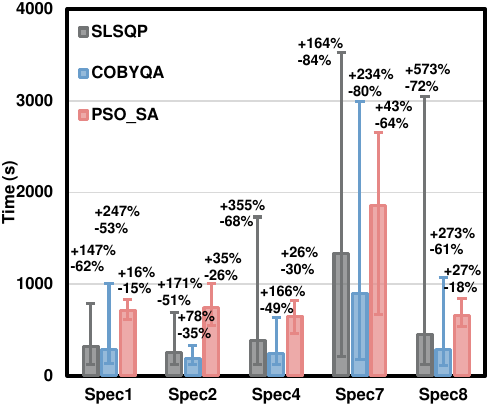}
  \caption{Variance of different inner solvers}
  \label{fig:R4C2}
\end{figure}

This section compares the PSO/SA implementation with deterministic solvers (SLSQP and COBYQA) by replacing the inner subproblem solvers while retaining the objective functions, parameter bounds, component-value conversion rules, and circuit acceptance criterion. Each method is evaluated over 20 trials on each of $S_1$–$S_8$, using a one-hour wall-clock limit per trial and a current allocation of [0.4, 0.3, 0.3]. Table~\ref{tabc} reports the success counts and average times. Fig.\ref{fig:R4C2} shows the variances of all three methods on tasks whose success rates exceed 50\%.

\begin{table}[!t]
\centering
\footnotesize
\caption{Comparison of inner solvers on $S_1$–$S_8$.}
\label{tabc}
\begin{tabular}{|c|c|c|c|c|c|c|}
\hline
\textbf{Spec} & \multicolumn{2}{c|}{\makecell{\textbf{PSO+SA}}} & \multicolumn{2}{c|}{\makecell{\textbf{SLSQP}}} & \multicolumn{2}{c|}{\makecell{\textbf{COBYQA}}} \\
\cline{2-7}
 & \makecell{pass\\@1} & \makecell{Avg\\Time(s)\\(\# Sim)} &  \makecell{pass\\@1} & \makecell{Avg\\Time(s)\\(\# Sim)} &  \makecell{pass\\@1} & \makecell{Avg\\Time(s)\\(\# Sim)} \\
\hline
$S_1$ & \textbf{20/20} & 715 & 19/20 & 322 & 18/20 & \textbf{291} \\
\hline
$S_2$ & \textbf{20/20} & 744 & 12/20 & 254 & 12/20 & \textbf{188} \\
\hline
$S_3$ & \textbf{5/20} & 1755 & 1/20 & 913 & 1/20 & \textbf{790} \\
\hline
$S_4$ & \textbf{20/20} & 653 & 14/20 & 381 & 17/20 & \textbf{239} \\
\hline
$S_5$ & \textbf{11/20} & 1127 & 7/20 & 977 & 9/20 & \textbf{235} \\
\hline
$S_6$ & \textbf{5/20} & 1586 & 3/20 & 1146 & 3/20 & \textbf{767} \\
\hline
$S_7$ & \textbf{18/20} & 1854 & 14/20 & 1335 & 11/20 & \textbf{895} \\
\hline
$S_8$ & 16/20 & 662 & 12/20 & 452 & \textbf{17/20} & \textbf{287} \\
\hline
\end{tabular}
\end{table}

\end{document}